\documentclass[a4paper,fleqn]{cas-dc}

\usepackage[authoryear]{natbib}
\def\tsc#1{\csdef{#1}{\textsc{\lowercase{#1}}\xspace}}
\tsc{WGM}
\tsc{QE}
\tsc{EP}
\tsc{PMS}
\tsc{BEC}
\tsc{DE}
\usepackage{graphicx}
\usepackage{caption}
\usepackage{subcaption}
\usepackage{color,soul}
\usepackage{multirow}
\usepackage{flushend}
\usepackage{array}
\usepackage{enumitem}
\usepackage{adjustbox}

\usepackage{pifont}
\newcommand{\cmark}{\ding{51}}%
\usepackage{algorithm}
\usepackage{algpseudocodex}
\algnewcommand\algorithmicinput{\textbf{Input:}}
\algnewcommand\Input{\item[\algorithmicinput]}
\usepackage{makecell}

\newcommand{\mysystem}{{CoQEx}}

\renewcommand{\paragraph}[1]{\medskip\noindent\textbf{#1.\mbox{\ \ }}}

\newcommand{\subscriptscore}[1]{$_{(#1)}$}

\begin{document}
\let\WriteBookmarks\relax
\def\floatpagepagefraction{1}
\def\textpagefraction{.001}

\shorttitle{Answering Count Questions with Structured Answers from Text}

\shortauthors{Ghosh et~al.}

\title [mode = title]{Answering Count Questions with Structured Answers from Text}                      
% \tnotemark[1,2]

% \tnotetext[1]{This document is the results of the research
%   project funded by the National Science Foundation.}

% \tnotetext[2]{The second title footnote which is a longer text matter
%   to fill through the whole text width and overflow into
%   another line in the footnotes area of the first page.}

\author[1,2]{Shrestha Ghosh}
% [type=editor,
% auid=000,bioid=1,
% prefix=Sir,
% role=Researcher,
% orcid=0000-0001-7511-2910]

\cormark[1]


\ead{ghoshs@mpi-inf.mpg.de}

% \ead[url]{www.cvr.cc, cvr@sayahna.org}

% \credit{Conceptualization of this study, Methodology, Software}

% \affiliation[1]{organization={Max Planck Institute for Informatics},
%     addressline={Saarland Informatics Campus}, 
%     city={Saarbruecken},
%     % citysep={}, % Uncomment if no comma needed between city and postcode
%     postcode={66125}, 
%     % state={},
%     country={Germany}}
\address[1]{{Max Planck Institute for Informatics},
    {Saarland Informatics Campus}, 
    {Saarbruecken},
    {66125}, 
    {Germany}}
% \affiliation[2]{organization={Saarland University},
%     addressline={Saarland Informatics Campus}, 
%     city={Saarbruecken},
%     % citysep={}, % Uncomment if no comma needed between city and postcode
%     postcode={66125}, 
%     % state={},
%     country={Germany}}
\address[2]{{Saarland University},
    {Saarland Informatics Campus}, 
    {Saarbruecken},
    {66125}, 
    {Germany}}

\author[1]{Simon Razniewski}
\ead{srazniew@mpi-inf.mpg.de}

\author[1]{Gerhard Weikum}
% [%
%   role=Co-ordinator,
%   suffix=Jr,
\ead{weikum@mpi-inf.mpg.de}
% \ead[URL]{www.sayahna.org}

% \credit{Data curation, Writing - Original draft preparation}

\cortext[cor1]{Corresponding author}
% \cortext[cor2]{Principal corresponding author}

% \fntext[fn1]{This is the first author footnote. but is common to third
% \fntext[fn2]{Another author footnote, this is a very long footnote and
%   it should be a really long footnote. But this footnote is not yet
%   sufficiently long enough to make two lines of footnote text.}

% % For a title note without a number/mark
% \nonumnote{This note has no numbers. In this work we demonstrate $a_b$
%   the formation Y\_1 of a new type of polariton on the interface
%   between a cuprous oxide slab and a polystyrene micro-sphere placed

\begin{abstract}
In this work we address the challenging case of answering count queries in web search, 
such as \textit{``number of songs by John Lennon''}. 
Prior methods merely answer these with a single, and sometimes puzzling number
or return a ranked list of text snippets with different numbers. 
This paper proposes a methodology for answering count queries with inference, contextualization and explanatory evidence. Unlike previous systems, our method infers final answers from multiple observations, supports semantic qualifiers for the counts, and provides evidence by enumerating representative instances. 
%In our previous work we proposed a methodology for answering count queries with inference, contextualization and explanatory evidence, which unlike previous systems, infers final answers from multiple observations, supports semantic qualifiers for the counts, and provides evidence by enumerating representative instances.
% Given the relatively new research topic we extend our previous work with extensive analysis of the characteristics of count queries in real user logs, and the challenges these pose to both knowledge-base question answering systems, and single-answer span text models.
Experiments with a wide variety of queries, including existing benchmark show the benefits of our method, and the influence of specific parameter settings. 
Our code, data and an interactive system demonstration are publicly available at {\small\url{https://github.com/ghoshs/CoQEx} and 
\url{https://nlcounqer.mpi-inf.mpg.de/}.}

% To promote further research on this underexplored topic, we release an annotated dataset of 5k queries with 200k relevant text spans.
% We extend research in this topic with a deep-dive into our annotated dataset CoQuAD and how the our proposed system {\mysystem} performs under different query conditions.  
% A challenging case in web search and question answering are count queries, such as “number of songs by John Lennon”. Prior methods merely answer these with a single, and sometimes puzzling number or return a ranked list of text snippets with different numbers. 
% Experiments with a wide variety of queries show the benefits of our method, and the influence of specific parameter settings. To promote further research on this underexplored topic, we release an annotated dataset of 5k queries with 200k relevant text span

% Experiments with a wide variety of queries show the benefits of our method.
% To promote further research on this underexplored topic, we release an annotated dataset of 5k queries with 200k relevant text spans.
\end{abstract}

% \includegraphics{figs/grabs.pdf}

% \begin{highlights}
% \item Research highlights item 1
% \item Research highlights item 2
% \item Research highlights item 3
% \end{highlights}

% Keywords
\begin{keywords}
% quadrupole exciton \sep polariton \sep \WGM \sep \BEC
Question Answering \sep Count Queries \sep Explainable AI
\end{keywords}

\maketitle

\section{Introduction}\label{sec:introduction}

\paragraph{Motivation and Problem} 
Question answering (QA)
%(QA) and entity search with telegraphic queries over web contents
and web search with telegraphic queries
have been greatly advanced over the last decade \citep{balog2018entity,diefenbach2018core,DBLP:journals/access/HuangXHWQFZPW20,DBLP:journals/semweb/UsbeckRHCHNDU19}.
Nevertheless, queries that can have multiple correct answers due to variance in semantic qualifiers (``top 10 albums'', ``singles albums'', ``remastered albums'') and alternative representations through instances remain underexplored and pose open challenges.
% Nevertheless, specific classes of queries are underexplored and pose open challenges.
This paper addresses the class of \textit{count queries},
to return the number of instances that have a certain property. Examples 
are:

\begin{itemize}
    \item {\small\textit{How many songs did John Lennon write for the Beatles?}}
    % \item \textit{How many people won the Fifa world cup both as player and as manager?}
    \item \small\textit{How many languages are spoken in Indonesia?}
    \item \small\textit{How many unicorn companies are there?}
\end{itemize}
% \GW{I suggest adding a third example of a different nature, ideally on something that is business or product oriented, to underline that these are frequent information needs for search engines}\\

\begin{figure*}[t]
    \centering
    \includegraphics[width=0.9\textwidth,trim={0cm 0cm 0.1cm 0cm},clip]{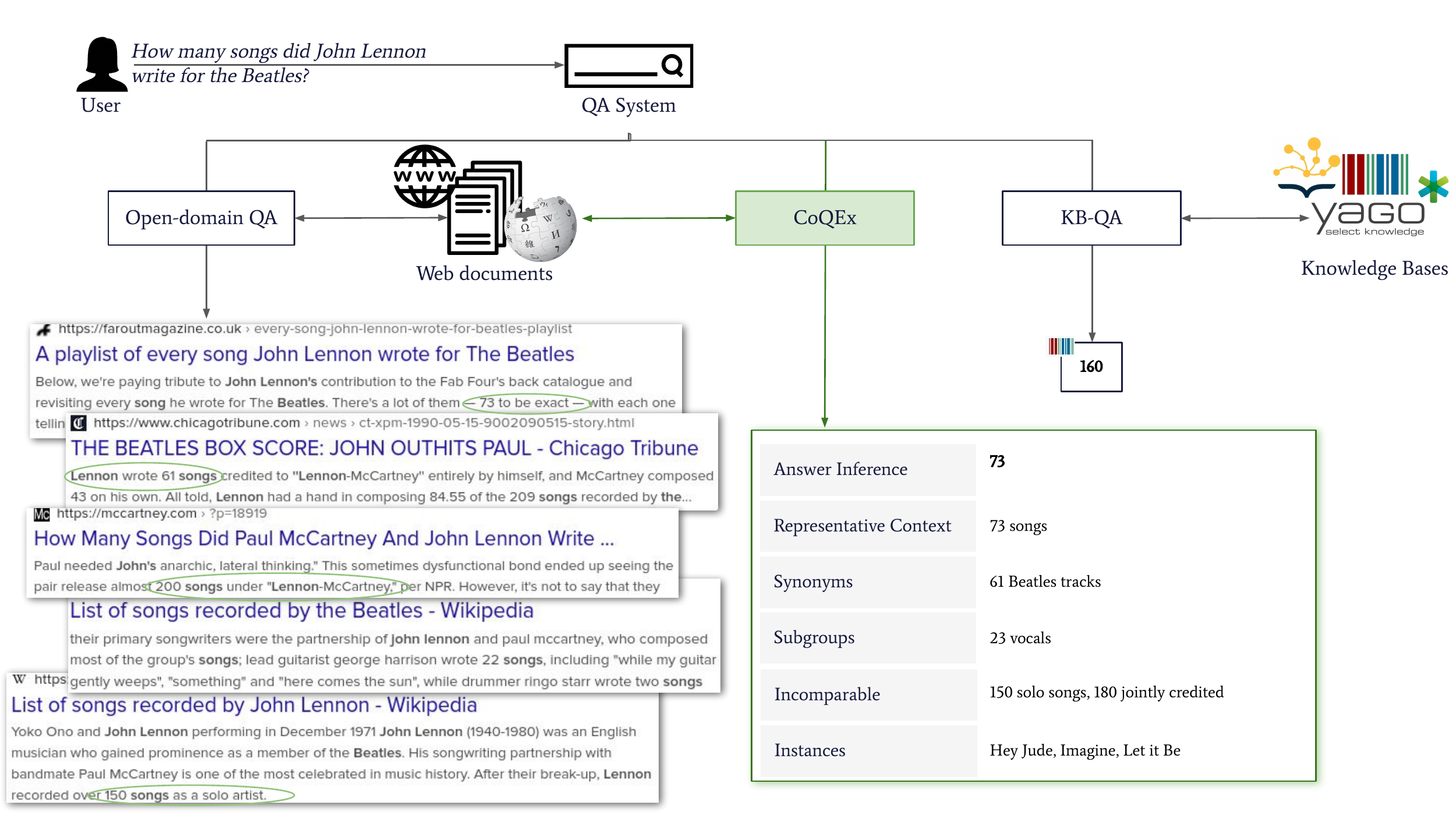}
    \caption{User experience with the state-of-the-art and proposed methodology for count question answering.}
    \label{fig:user_interation_sota}
\end{figure*}

Count queries are frequent in search engine logs as well as QA benchmarks~\citep{rajpurkar2016squad,kwiatkowski2019natural,dubey2019lc,voorhees2001overview}.
If the required data is in a structured knowledge base (KB) such as Wikidata~\citep{vrandevcic2014wikidata}, then answering is relatively straightforward.
However, KBs are limited not only by their sparsity, but also by the lack of direct links between instances and explicit counts when both are present~\citep{ghosh2020uncovering}. 
% on this aspect and miss out often.
% For example, although Wikidata contains 
% some of 
%McCartney's 217
% (as on {09.12.2021}) 217 of Lennon's
% compositions, it does not have full coverage and no explicit count either. 
Besides, evaluating the additional condition \textit{``for the Beatles''} (i.e., a subset of his songs) is beyond their scope. 
% \GW{added the SE angle early on, as WSDM reviewers said SE would do this already}\\
In Figure~\ref{fig:user_interation_sota}, \textit{160} is the output to a SPARQL query which counts the number of songs by Lennon performed by The Beatles\footnote{\url{https://w.wiki/4XVq}}. The query translation, here performed by an expert user, is a challenge in itself, which is beyond the scope of this work. Nevertheless, for a user who is unaware of the composer duo Lennon-McCartney, would hardly doubt the output of 160 songs to the SPARQL query, which contains songs jointly written by Lennon with his co-band member. 

Search engines are the commercial state-of-the-art that are exposed to real-life user queries. They handle popular cases reasonably well, but also
fail on semantically refined requests (e.g., \textit{``for the Beatles''}), merely returning either a number without explanatory evidence or multiple candidate answers with high variance (Figure~\ref{fig:user_interation_sota}). We also see incorrect spans being highlighted \textit{``22 songs''} in the fourth snippet, which is actually a count of songs written by George Harrison.
% Search engines return matches 
% on 241 websites, 
% with counts such as \textit{``over 150 songs''} and \textit{``73''} when queried for \textit{``how many songs by John Lennon''}. With the additional semantic qualifier \textit{``for the Beatles''} the counts \textit{``209''}, \textit{``73''} and \textit{``61''} come up in the search result snippets (as on 09.12.2021 from Google and Bing). 
% The situation gets worse for less popular topics.
% , which is often dissatisfying to users. 
% and semi-structured contents (e.g., lists) 
% \noindent 

% We tackle these limitations by tapping into
% textual web contents.
Answering count queries from web contents thus poses several challenges:

\begin{itemize}
    \item[1.] \textit{Aggregation and inference:} 
    % Web search often yields multiple candidate answers, sometimes with high variance. For example, for the wildcard pattern query ``Lennon wrote * songs'', major search engines return matches on 241 websites, with counts such as ``over 200'', ``61'' etc.
    % \GW{how do you run a wildcard query on Google ???}\\
    % \GW{why 241 matches? shouldn't there be many thousands?}\\
    % , ``hundreds''. 
    %Outputting the answer only from a single website is a lucky draw, and risks neglecting corroborating evidence from other candidates.
    Returning just a single number from the highest-ranked page can easily go wrong.
    Instead, joint inference over a set of candidates, with an awareness of the distribution and other signals, is necessary for a high-confidence answer.
    \item[2.] \textit{Contextualization:} Counts in texts often 
    come with contexts on the relevant instance set.
    % , such as ``in high school'', or ``for the Beatles''. 
    For example, John Lennon co-wrote about 180 songs 
    %together with Paul McCartney, 
    for the Beatles,
    % 43 credited to Lennon-McCartney on his own, 
    150 as a solo artist, etc.
    For correct answers it is crucial
    to 
    %go beyond IR-style context similarity, and to explicitly extract and process these semantic qualifiers.
    capture context from the underlying web pages
    and properly evaluate these kinds of semantic qualifiers.
    \item[3.] {\textit{Explanatory Evidence:}} 
    A numeric answer alone, such as \textit{``180''} for the Beatles songs by Lennon, is often unsatisfactory. 
    The user may even perceive this as non-credible, and think that it is too high as she may have only popular songs in mind.
    %thinking only in terms of the solo authored ones, which are much less. 
    It is, therefore, crucial 
    % provide user-comprehensible explanations, 
    %{explain these counts,}
    %by showing notable instances.
    to provide users with explanatory evidence.
    % and web evidence for them.
\end{itemize}
\paragraph{Contribution}
This paper presents {\mysystem}, \underline{Co}unt \underline{Q}uestion answering with \underline{Ex}planatory evidence, which answers count queries via three components: i) \textit{answer inference} ii) \textit{answer contextualization} and, iii) \textit{answer explanation}.

Given a full-fledged question or telegraphic query and relevant text passages, {\mysystem} 
% retrieves a set of candidate results and
applies joint inference to compute a high-confidence answer for the count itself.
It provides contextualization of the returned count answer,
through semantic qualifiers into equivalent or subclass categories, and extracts a set of representative instances as explanatory evidence, exemplifying the returned number for
enhanced credibility and user comprehension.
%retrieves a ranked-list of instances that explain the counts.
%
% , by using ranking techniques to identify informative instances.
% The results from both tasks are combined to boost the accuracy of each other. 

\noindent Contributions of this work are:

\begin{enumerate}
    \item[1.] introducing the problem of 
    count query answering with explanatory evidence;
    \item[2.] developing a 
    method
    %module 
    for inferring high-confidence counts from noisy candidate sets;
    \item[3.] developing 
    %two modules 
    techniques
    to provide answer contextualization and explanations;
    % from count-modified noun phrases, %%SG: I'm refraining from using the term CNPs until introduced in CoRe in the methodology.
    \item[4.] evaluating 
    % Co$^3$ 
    {\mysystem}
    against state-of-the-art baselines 
    %and diverse query datasets, and; 
    on a variety of test queries;
    % and
    \item[5.] releasing an annotated data resource with 5k count queries and 200k text passages,
    % with annotations,
    available at {\small\url{https://github.com/ghoshs/CoQEx}}.
    % {\small\url{https://tinyurl.com/countqueryappendix}}.
\end{enumerate}

\paragraph{Previous publication}
The present manuscript substantially extends a short paper by~\cite{ghosh2022answering} by a thorough analysis of the problem of count question answering. In particular, we analyse the characteristic of real count queries, and dissect the difficulty and tractability of each of the subproblems. The major extensions to the short paper are:
\begin{enumerate}
    \item We substantially expanded the task data, and, based on novel annotation, provide an in-depth analysis of count query characteristics in Section 5.2. In particular, we group queries by complexity (in reach for today's structured approaches, in reach for textual approaches, out of reach), and analyze baseline and system performance for each of them (Tables 3 and 5 in 7.2 and 7.5). We also analyze the domains and stability properties of count queries (Figures 3 and 4), and added another related dataset (NaturalQuestions) to corroborate the findings from our newly-built CoQuAD dataset. Our findings include that our proposed system is resilient to harder query types, with comparable precision at moderate loss of coverage/recall.
    \item We provide a novel analysis of count contextualization across different QA paradigms in Section 7.3, finding that KB-QA and search engines struggle with this for principled and/or pragmatic reasons, thus further motivating our answer contextualization step.
    \item We provide a full component analysis on our CoQEx system in Section 8, specifically through Table 8 and Figures 5, 6 and 7. A main takeaway is how significant thresholding choices are in answer inference and answer explanation, compared with a lower impact of the pre-training choice and consolidation strategy. For answer contextualization, we observe that synonyms are easier to discern than subgroups and incomparables.
    \item We provide provide a discussion in Section 9 which highlights open challenges for future work, specifically, how count contexts and instance explanations coexist and contribute to better user comprehension. We also examine insightful case studies which show the potential of CoQEx in explaining count queries through hierarchical count contexts and instances.
\end{enumerate}

\section{Related Work}

Where structured data is available in KBs, structured QA is the method of choice for count question answering, and previous work of~\citet{ghosh2020uncovering} has looked at identifying count information inside KBs. 
 However, for many topics, no relevant count information can be found in KBs. 
For example, Wikidata contains 217 songs attributed to John Lennon\footnote{\url{https://w.wiki/4XVq}},
% https://w.wiki/5E4e -- songs composed by Lennon, performed by the Beatles
% \GW{ToDo: check the number!!!}, 
but is incomplete in indicating whether these written for the Beatles or otherwise.
In the KB-QA domain, systems like QAnswer developed by~\cite{diefenbach2019qanswer} tackle count queries
by aggregating instances using the SPARQL \texttt{count} modifier. This is liable to incorrect answers, when instance relations are incomplete.
Attempts have also been made to improve recall by hybrid QA over text and KB, yet without specific consideration of counts~\citep{lu2019answering,xu2016hybrid}.

% \paragraph{Count Information in QA} 
% In~\cite{enriching2018mirza}, it is reported that 5\%-10\% of queries in popular QA datasets are count queries. 
State-of-the-art systems typically approach QA via the machine reading paradigm~\citep{karpukhin2020dense,joshi2020SpanBERT,sanh2019distilbert,chen2017reading,dua2019drop}, where the systems find the best answer in a given passage. The retriever-reader approach in open-domain QA uses several text segments 
% (multi-paragraphs from a single or multiple documents) 
to return either a single best answer~\citep{chen2017reading,wang2018evidence} or a ranked list of documents with the best answer per document~\citep{karpukhin2020dense}.
The DPR system by~\cite{karpukhin2020dense}\footnote{\url{http://qa.cs.washington.edu:2020}} returns 
\textit{``approximately 180''} from its rank-1 text passage
to both, the simple John Lennon query, and the refined variant with \textit{``\dots for the Beatles''}.
The other top-10 snippets include false results such as \textit{``five''} and contradictory information such as \textit{``180 jointly credited''}
% Lennon and McCartney together, 
(as if Lennon had not written any songs alone).
Thus, QA systems are not robust (yet) and lack explanatory evidence beyond merely returning the top-ranked text snippet. 

% \GW{need a short par on SEs here, proposed text follows:}\\
Attempts have also been made to improve recall by hybrid QA over text and KB, yet without specific consideration of counts~\citep{lu2019answering,xu2016hybrid,saha2020question}.
Search engines can answer simple count queries from their underlying KBs, if present, a trait which we exploit to create our CoQuAD dataset (Section~\ref{sec:dataset}). But more often they return informative text snippets, similar to QA-over-text systems.
The basic Lennon query has a highest-ranked Google snippet with
\textit{``more than 150''} when given the telegraphic input
\textit{``number of songs by John Lennon''} and
\textit{``almost 200''} when given the full-fledged question
\textit{``how many songs did John Lennon write''}. For the latter case,
the top-ranked snippet talks about the composer duo
\textit{``John Lennon and Paul McCartney''}. 
When refining the query by qualifiers, this already puzzling
situation becomes even more complex with \textit{``84.55 of 209 songs''} being ranked first followed by varying counts such as \textit{``18 Beatles songs''} 
(co-written with McCartney) 
and \textit{``61''} 
(written separately).  
Because of the lack of consolidation, the onus is on the user to decide whether there are multiple correct answers across text segments. 

% \paragraph{QA Datasets}
In~\cite{enriching2018mirza}, it is reported that 5\%-10\% of queries in popular QA datasets are count queries.  Current text-based QA systems and datasets largely ignore consolidation over multiple documents, since the target is to produce a single answer span or document. QA systems are tested on reading comprehension datatsets, the most popular being SQuAD~\citep{rajpurkar2016squad}, CoQA~\citep{reddy2019coqa} and more recent being DROP~\citep{dua2019drop}, on open domain QA datatsets such as Natural Questions~\citep{kwiatkowski2019natural}, TriviaQA~\citep{joshi2017triviaqa} and on KBs with datasets such as LC-QuAD 2.0~\citep{dubey2019lc}, WebQuestions~\citep{berant2013semantic} and QALD \cite{qald}. Reading comprehension and open domain QA datasets are annotated with answer spans,  while the datasets for KB-QA come with a single or a list of answers, but no further context. Multi-document-multi-hop reasoning datasets, in turn, focus on chaining evidence~\citep{dua2019drop,bauer2018commonsense}. 

The evaluation metrics for reading-comprehension style benchmarks typically employ strict matching requirements, like measuring accuracy, F1-score and exact match~\cite{zeng2020survey}. On a question level, these metrics measure the token-level overlap, which does not transfer well to count queries, especially counts which do not have one authoritative answer and are an inference over multiple documents. We propose relaxed metrics for evaluation in Section~\ref{sec:experimental_setup}.

% \paragraph{Long Form Question Answering}
It is recognized that just literally answering questions is often not sufficient for use cases. One line of work,~\cite{kacupaj2020vquanda}, tackles this by returning comprehensive answer in full sentences, using templates. Another line,~\cite{krishna2021hurdles}, concerns long form question answering, where the QA model retrieves multiple relevant documents to generate a whole answer paragraph. The ELI5 dataset by~\cite{fan2019eli5} contains diverse open-ended queries with supporting information from relevant web sources. While the setting is related, long form QA is concerned with generating textual answers evidenced on multiple documents, while we focus on answering count queries by consolidating counts and grounding them in instances.    

\section{Design Space and Architecture}

%\paragraph{Existing Paradigms}
Traditional open domain QA architectures involve query analysis, document retrieval and answer extraction, where named-entity recognition (NER) is an important component for recognizing named entities of the answer type~\citep{zhu2021retrieving}, with the Text REtrieval Conference (TREC) QA tracks leading the research in fact-based question answering~\citep{voorhees2001overview}. 
% While traditional open domain QA architectures rely heavily on named-entity recognition (NER)~\citep{zhu2021retrieving} with the Text Retrieval Conference (TREC) QA tracks leading the research in fact-based QA~\citep{voorhees2001overview},
Current research employs deep learning with the benefit of achieving end-to-end trainable systems. In this direction we discussed open-domain QA systems in the reader-retriever paradigm, KB-QA which translates natural language questions to structured queries finally executed over a KB to return the answer and the hybrid QA setting which uses a mix of structured KBs and texts to answer natural language questions.
% ~\citep{saha2020question}.

\begin{figure}[t]
    \centering
    \includegraphics[width=0.49\textwidth,trim={3.8cm 2.3cm 3.5cm 2.3cm},clip]{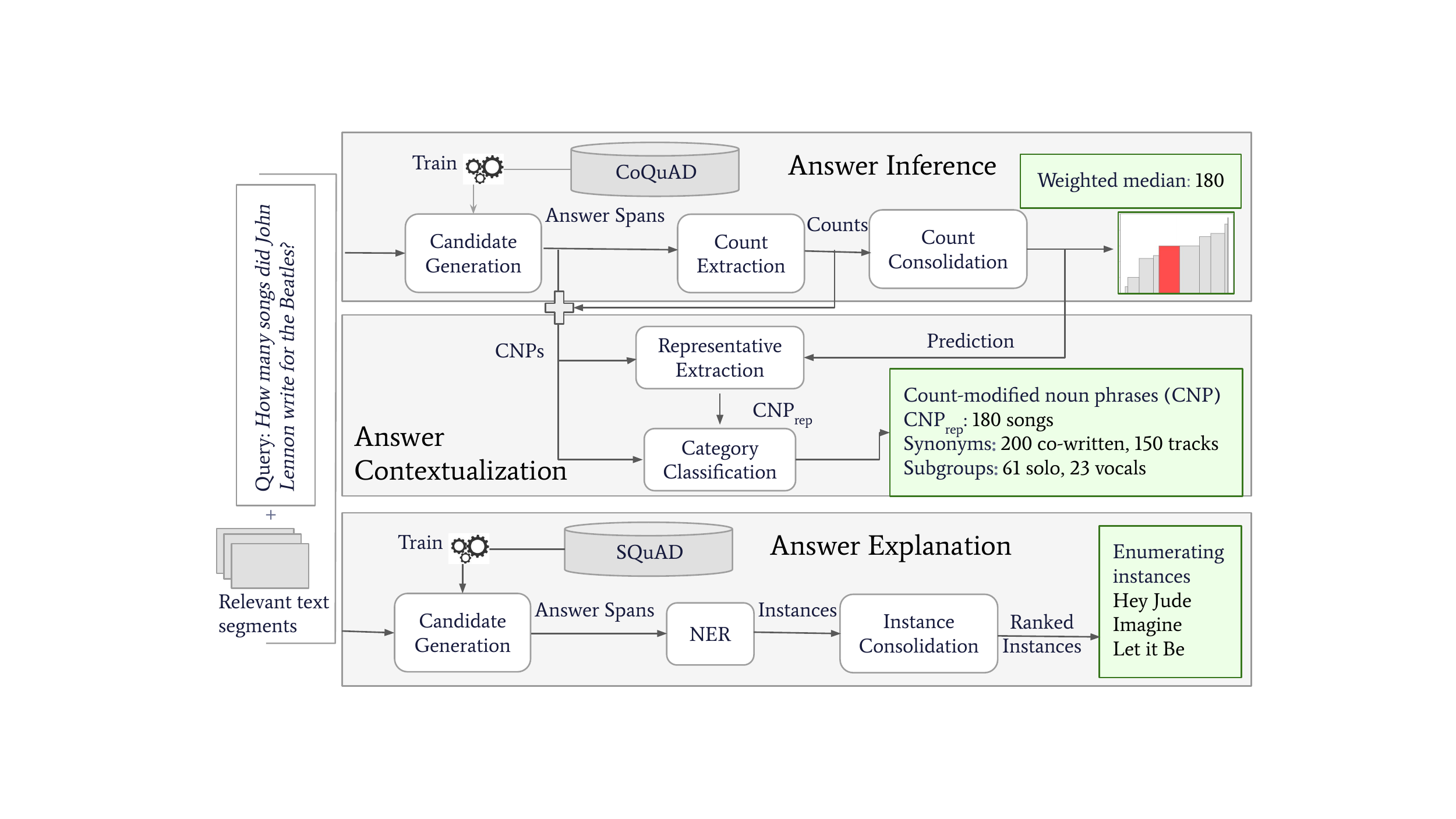}
    \caption{System overview of {\mysystem} from~\citet{ghosh2022answering}. 
    % \sr{Remove plural ``s'' from CoRe and CoEx names. Also, consider different color for output boxes?}
    % }
    }
    \label{fig:system_arch}
    
\end{figure}

We approach count query answering by a combination of per-document answer span prediction, context extraction, and consolidation of counts and instances across documents.
%For count queries, answer types are uniformly counts, and the mileage thus comes from recognizing counts in the right context, and consolidating corroborating candidate answers into a most likely value.
% We therefore approach the task of count QA over text by utilizing evidence from multiple document sources for solving the tasks of \textit{answer inference}, \textit{answer contextualization} and 
% \textit{count explanation}
Figure~\ref{fig:system_arch} gives the overview of {\mysystem}. We consider as input a query that asks for the count of named entities that stand in relation with a subject, for instance full queries like \textit{``How many songs did John Lennon write for the Beatles''}, or a keyword query like \textit{``songs by lennon''}.

We further assume that relevant documents or passages are given. This could be the result of a standard keyword/neural embedding-based IR procedure over a larger (locally indexed) background corpus, like Wiki\-pedia or the Web. We explain the methodology of {\mysystem} in the next section.

\section{Methodology}\label{sec:methodology}

% We approach count question answering by a combination of per-document answer span prediction, context extraction, and consolidation of counts and instances across documents.
% %For count queries, answer types are uniformly counts, and the mileage thus comes from recognizing counts in the right context, and consolidating corroborating candidate answers into a most likely value.
% % We therefore approach the task of count QA over text by utilizing evidence from multiple document sources for solving the tasks of \textit{answer inference}, \textit{answer contextualization} and 
% % \textit{count explanation}
% Fig.~\ref{fig:system_arch} gives the overview of {\mysystem}. We consider as input a query that asks for the count of named entities that stand in relation with a subject, for instance full queries like \textit{``How many songs did John Lennon write for the Beatles''}, or a keyword query like \textit{``songs by lennon''}.

% We further assume that relevant documents or passages are given. This could be the result of a standard keyword/neural embedding-based IR procedure over a larger (locally indexed) background corpus, like Wiki\-pedia or the Web. % Yet this is merely a pragmatic choice, locally indexed text corpora could be plugged in as well.
{\mysystem} extracts counts and instances (entity-mentions) from the text segments to subsequently i) consolidate the counts to present the best answer, ii) present 
contextualization
as a means to semantically qualifying the predicted count, and iii) ground the count in instances. We denote the set of relevant text segments for a query by $D$.

\subsection{Answer Inference}\label{subsec:coin}
In order to generate count candidates, we use the popular SpanBERT model from~\cite{joshi2020SpanBERT}, trained on the CoQuAD train split.
Span prediction models 
return general text spans, which may contain worded answers (\textit{``five children''}, $\textit{Conf}=0.8$), modifier words and other context 
(\textit{``17 regional languages''}, $\textit{Conf}=0.75$), where \textit{Conf} is the confidence score of the model. 
These answer spans have two components - the count itself and qualifiers, which we separate with the help of fixed rules and the CogComp Quantifier by~\citep{roy2015reasoning}. 

% The Quantifier works well on phrases, but does not guarantee extraction if the span is just a number (\textit{17}) or its worded form (\textit{seventeen}) with no additional context. We therefore modify the count extraction to handle this case. 
% 
Algorithm~\ref{algo:inference} shows the outline for answer inference. We run all relevant documents, $D$ for a given query though the span prediction model to get the candidate spans comprised of the answer span, $c.\textit{Span}$, and model confidence, $c.\textit{Conf}$, (\texttt{Line 3-4}). If the span is non-empty and confidence of the model is higher than a threshold, $\theta$, then we extract integer count from the span using the \texttt{ExtractCount} function (\texttt{Line 5-6}). If the integer extraction in non-empty, we save the span, the extracted integer and the model confidence (\texttt{Lines 7-9}). 

The \textsc{ExtractCount()} sequentially applies rule-specific conversions before applying CogComp Quantifier to achieve maximum recall. The function first applying type conversion (\texttt{int}(\textit{``17''})$\rightarrow$\textit{17}, \texttt{float}(\textit{``17.0''})$\rightarrow$\textit{17.0}), followed by a dictionary based look-up for worded to integer conversion (\textit{``seventeen''}$\rightarrow$\textit{17}) and lastly the CogComp Quantifier, proceeding only when the previous conversion yields empty results. The counts are further cleaned by removing fractions $\in (0,1)$, since counts are whole numbers.

If there is at least one count extracted from the relevant document set, we consolidate the counts using either of the proposed consolidation methods defined in the \texttt{Consolidate} function (\texttt{Lines 10-13}), else the answer inference is empty (\texttt{Line 15}). 

To consolidate the resulting candidate counts into a prediction $C_\textit{pred}$, we compare four methods:
% \begin{enumerate}
\begin{itemize}
    \item[1.] \textit{Most confident:} The candidate given the highest confidence by the neural model. This is commonly used in textual QA~\citep{chen2017reading,wang2018evidence}.
    % \sr{Cite what?}.
    \item[2.] \textit{Most frequent:} A natural alternative is to rank answers by frequency, and prefer the ones returned most often. 
    % \sr{TODO: Add in experiments}
\end{itemize}

While \textit{most confident} may be susceptible to single outliers, \textit{most frequent} breaks down in cases where there are few answer candidates. But unlike textual answers, numbers allow further statistical aggregation:
% \begin{enumerate}[resume]
\begin{itemize}
    \item[3.] \textit{Median:} The midpoint in the ordered list of candidates.
    \item[4.] \textit{Weighted Median:} The median can be further adapted by weighing each candidate with the model's score.
\end{itemize}

For example, for the candidate set $\{150_{0.9}, 160_{0.8}, 180_{0.4},\allowbreak 180_{0.4}, 210_{0.3}\}$ (confidences as subscripts), most confident would output 150, most frequent and median would return 180, and the weighted median 160.

% We evaluate whether such number-specific aggregations add value over standard QA answer selection in Section~\ref{sec:experiments}.

\begin{algorithm}
\caption{Extracting answer inference}\label{algo:inference}
% \hspace*{\algorithmicindent} 
\textbf{Input:} Count query, $q$,\\
\hspace*{1cm}set of relevant text segments, $D$,\\
\hspace*{1cm}span prediction model, $\textsc{SpanPrediction}$,\\
\hspace*{1cm}span selection threshold, $\theta$,\\
\hspace*{1cm}count extraction function, $\textsc{ExtractCount}$,\\
\hspace*{1cm}consolidation function, $\textsc{Consolidate}$.\\
% \hspace*{\algorithmicindent} 
\textbf{Output:} Answer Inference, $C_\textit{pred}$,\\
\hspace*{1cm}List of count span and extracted integer tuples, $C$.
\begin{algorithmic}[1]
    \State $C \gets \{\}$ \Comment{Passed to contextualization}
    \State $\textit{WeightedC} \gets \{\}$ \Comment{Counts with confidence}
    \For{$d \in D$}
        \State $c \gets \Call{SpanPrediction}{d, q}$
        \If{$c.\textit{Span} \neq \textit{None and }c.\textit{Conf} > \theta$} 
            \State $i \gets \Call{ExtractCount}{c.\textit{Span}}$
            \If{$i \neq \textit{None}$}
                \State $\textit{C} \gets \textit{C} \cup (c,i)$ 
                \State $\textit{WeightedC} \gets \textit{WeightedC} \cup (i, c.\textit{Span})$
            \EndIf
        \EndIf
    \EndFor
    \If{$\textit{WeightedC} \neq \{\}$}
        \State $\textit{WeightedC} \gets \Call{SortAscending}{\textit{WeightedC}}$
        \LComment{Return the weighted median of the counts.}
        \State $C_\textit{pred} \gets \Call{Consolidate}{\textit{WeightedC}}$ 
    \Else
        \State $C_\textit{pred} \gets Null$
    \EndIf
    
    \State \Return $C_\textit{pred}, C$
\end{algorithmic}
\end{algorithm}

\subsection{Answer Contextualization}
% \subsubsection{\bf Count-modified Noun Phrases}
The answer candidates from the previous module often contain nouns with phrasal modifiers, such as \textit{17 regional languages}. %The answer span, count and score tuples we obtained in CoIn are re-used in CoRe. Not all answer spans contain counts (``few languages'', $\textit{count}=0$, $\textit{score}=0.5$). To those that do contain them we define as 
We call these 
\textit{count-modified noun phrases} (CNPs). These CNPs stand in some relation with the predicted count from the answer inference module as explained in Algorithm~\ref{algo:contextualization}. 
The representative CNP, CNP$_\text{rep}$,  
% As representative CNP, we use 
which best accompanies the predicted count is first chosen and then compared with the remaining CNPs. Since answer inference uses a consolidation strategy, we select the CNP with count equal to $C_\textit{pred}$ having the highest confidence as CNP$_\text{rep}$ (\texttt{Line 2}).

The remaining CNPs are categorized as follows:
% By comparing count-modified noun phrases (CNPs) based on their relation with the representative CNP of the predicted answer (CNP$_\textit{rep}$), we can discern their semantics:
\begin{itemize}
    \item[1.] \textit{Synonyms}: CNPs, whose meaning is highly similar to CNP$_\textit{rep}$ and accompanying count is within a specified threshold, $\alpha$, of the predicted count (\texttt{Lines 6-7}), where $\alpha$ is between 0\% and 100\%, 0 being most restrictive.. 
    % E.g., \textit{tongue} is predicted as synonym of \textit{language}, if it occurs with counts $\langle 530, 810, 600\rangle$.
    \item[2.] \textit{Subgroups}: CNPs which are semantically more specific than CNP$_\textit{rep}$, and are expected to count only a subset of the instances counted by CNP$_\textit{rep}$, such that the accompanying count is lower than the synonyms set (\texttt{Lines 8-9}). %  where the accompanying count is lower than the predicted answer. E.g., \textit{regional languages} is predicted as subgroup of \textit{language}, if it occurs with counts $\langle 23, 17, 42\rangle$ and the predicted query answer is 700.
    \item[3.] \textit{Incomparables}: CNPs which count instances of a completely different type indicated by negative cosine similarity (\texttt{Lines 4-5}) or an accompanying count higher than the synonyms (\texttt{Line 11}).
    %If a CNP is semantically dissimilar to CNP$_\textit{rep}$ or the accompanying count of CNP is much greater than the predicted count we categorize it as unrelated. E.g., \textit{speakers}, which occurs with counts in the order of millions, is semantically similar to \textit{language}, but does not relate to the language count.
\end{itemize}

\begin{algorithm}
\caption{CNP Category Classifier}\label{algo:contextualization}
% \hspace*{\algorithmicindent} 
\textbf{Input:} Answer Inference, $C_\textit{pred}$,\\
\hspace*{1cm}synonym threshold, $\alpha$,\\
\hspace*{1cm}list of count spans and extracted integer tuples, $C$ (from Algorithm~\ref{algo:inference})\\
% \hspace*{\algorithmicindent} 
\textbf{Output:} Representative CNP, $\textit{CNP}_\textit{rep}$,\\
\hspace*{1cm}List of CNP categories, $\textit{Categories}$
\begin{algorithmic}[1]
    \State $\textit{Synonyms, Subgroups, Incomparables} \gets \{\}, \{\}, \{\}$
    \State $\textit{CNP}_\textit{rep} \gets \underset{c}{\mathrm{argmax}} \{c.\textit{Conf} \mid i = C_\textit{pred}, (c, i) \in C \}$
    \For{$(c,i) \in C\setminus(\textit{CNP}_\textit{rep}, C_\textit{pred})$}
        % \State $\textit{CNP} \gets c.\textit{Span}$
        \If{$\Call{CosineSim}{c.\textit{Span}, \textit{CNP}_\textit{Rep}.\textit{Span}} <= 0$}
            \State $\textit{Incomparables} \gets \textit{Incomparables} \cup c$
        \ElsIf{$i \in C_\textit{pred}\pm \alpha $}
            \State $\textit{Synonyms} \gets \textit{Synonym} \cup c$
        \ElsIf{$i <  C_\textit{pred} - \alpha C_\textit{pred}$}
            \State $\textit{Subgroups} \gets \textit{Subgroups} \cup c$
        \Else
            \State $\textit{Incomparables} \gets \textit{Incomparables} \cup c$
        \EndIf    
    \EndFor
    \State \Return $\textit{CNP}_\textit{rep}, \textit{Synonyms}, \textit{Subgroups}, \textit{Incomparables}$
\end{algorithmic}
\end{algorithm}

\noindent
We assign these categories based on (textual) semantic relatedness of the phrasal modifier, and numeric proximity of the count.
For example, \textit{``regional languages''} is likely a subgroup of \textit{``700 languages''}, especially if it occurs with counts $\langle 23, 17, 42\rangle$.
\textit{``tongue''} is likely a synonym, especially if it occurs with counts $\langle 530, 810, 600\rangle$. \textit{``Speakers''} is most likely incomparable, especially if it co-occurs with counts in the millions.
% The representative CNP, CNP$_\text{rep}$,  
% % As representative CNP, we use 
% which best accompanies the predicted count is first chosen and then compared with the remaining CNPs. Sine CoIn prediction comes from a consolidation strategy, we select the CNP with count within $\pm$30\% of the CoIn prediction and highest score as the CNP$_\text{rep}$.
% a noun phrase that well define the count prediction highest confidence among all those within $\pm$30\% of the CoIn prediction. 
CNPs with embedding-cosine similarity~\citep{reimers2019sentencebert} less than zero are categorized as incomparable, while from the remainder, those with a count within $\pm\alpha$ are considered synonyms, lower count CNPs are categorized as subgroups, and higher count CNPs as incomparable. 

For instance, for the query \textit{``How many languages are spoken in Indonesia''}, with a prediction 700, \textit{estimated 700 languages} would be the $\textit{CNP}_\textit{rep}$, \{\textit{700 languages}, \textit{750 dialects}\} would be classified as synonyms, \{\textit{27 major regional languages}, \textit{5 official languages}\} as subgroups and \{\textit{2000 ethnic groups}, \textit{85 million native speakers}\} as incomparables (Table~\ref{tab:count_modified_np_example}).
% An illustrative example is provided in Table~\ref{tab:count_modified_np_example}.

\begin{table}[t]
    \centering
    \caption{CNPs with their categories for a query and confidence scores as subscripts.}
    \label{tab:count_modified_np_example}
    \begin{adjustbox}{width=0.49\textwidth}
    \begin{tabular}{|ll|}
    \hline
        {Query} & \textit{How many languages are spoken in Indonesia?}\\ 
    \hline
        {$\textit{CNP}_\textit{Rep}$} & estimated 700 languages\subscriptscore{0.8}\\ 
        {Synonyms} & 700 languages\subscriptscore{0.7}, about 750 dialects\subscriptscore{0.7}\\
        {Subgroups} & 27 major regional languages\subscriptscore{0.6}, 5 official languages\subscriptscore{0.8} \\
        {Incomparables} &  2000 ethnic groups\subscriptscore{0.4}, 85 million native speakers\subscriptscore{0.5}\\
    \hline
    \end{tabular}
    \end{adjustbox}
\end{table}

\subsection{Answer Explanation}\label{subsec:coex}
% \paragraph{Enumerating Instances}
Beyond classifying count answer contexts, showing relevant sample instances is an important step towards explainability. To this end, we aim to identify entities that are among the ones counted in the query using Algorithm~\ref{algo:explanation}.
% and to rank them by their relevance as described by Algorithm~\ref{algo:coex},
% We consider a QA-based consolidated strategy to identify those instances and compare against context frequency and type compatible frequency with further details on the latter two in Sec.~\ref{sec:experiments}.
% given the same query and the set of related text segments from input in Section~\ref{subsec:input}.
% , our aim is to generate a ranked list of unique instances as described by Alg.~\ref{algo:coex}. 
% The first task is to obtain the instances and for this 

Let 
% $D$ denote the set of text segments (documents or spans), and 
$I$ denote the inverted dictionary of instances where $I[i]$ contains the text IDs and confidence scores of the instance $i$. 
% We use NER to identify instances, either in whole contexts or in the spans returned by a QA model, and rank them by various features. 
We collect the answers from a QA model to create a more precision-oriented candidate space.
% a QA-based consolidated strategy,
% to identify the instances. 
We again use the SpanBERT model (fine-tuned on SQuAD 2.0 dataset) to obtain candidates (tuples comprising answer span, $c.\textit{Span}$, and the model confidence, $c.\textit{Conf}$) from every document (\texttt{Lines 4-5}), this time with a modified query, replacing \textit{``how many''} in the query with \textit{``which''} (or adding it), so as to not confuse the model on the answer type (\texttt{Line 1}). If the span is non-empty and has a confidence higher than threshold, $\theta$, we extract named entities from the span (\texttt{Lines 7-8}) using an off-the-shelf NER. We create an inverted index of these instances, keeping track of the text segment it belongs to and the span prediction (\texttt{Lines 9-10}). The instances are then scored globally using either of the following alternative consolidation scoring approaches defined in the \textsc{ConScore} function in \texttt{Lines 11-12}. The instances are then ranked in decreasing order of their consolidated scores (\texttt{Line 13}).

% We now derive ranked lists of these instances 
% using the following alternative approaches:
% \begin{enumerate}
The alternatives for instance consolidation are as follows. We normalize the consolidation scores for comparison across instances and strategies. All consolidation strategies lie between $[0,1]$. 
\begin{itemize}
\item[1.] \textit{QA w/o Consolidation.} In the spirit of conventional QA, where results come from a single document, we return instances from the document with the most confident answer span.

\item[2.] \textit{QA + Context Frequency.} 
% Here, we use the answers from a QA model to create a more precision-oriented candidate space.
% a QA-based consolidated strategy,
% to identify the instances. 
% An extractive QA model~\cite{joshi2020SpanBERT} answers a modified version of the count query, replacing "how many" in the query with "which", so as to not confuse the model on the answer type. NER then identifies instances in the answer spans. 
The instances are ranked by their frequency, $S[i] = \frac{|I[i]|}{|D|}$.

\item[3.] \textit{QA + Summed Confidence.} We rank the instances based on the summed confidence of all answer spans that contain them, %the parent answer span as returned by the QA model,
 $S[i] = \frac{{\sum_{(\cdot,c) \in I[i]} c.\text{Conf}}}{|I[i]|}$.

\item[4.] \textit{QA + Type Compatibility.} Here instances are ranked by their compatibility with the query's answer type, extracted via the dependency parse tree. %\footnote{as returned by \texttt{en\_core\_web\_sm} model in \url{https://spacy.io}}.
We obtain the answer type by extracting the first \texttt{noun} token and any of its preceding \texttt{adjectives} from the dependency parse tree of the query. 
% We then check the type compatibility of candidate instances via
We form a hypothesis \textit{``(instance) is a (answer type)''} and use the probability of its entailment from the parent sentence in the context from which the instance was extracted to measure type compatibility. We use~\cite{liu2019roberta} to obtain entailment scores, which are again summed over all containing answer spans, %(\textit{Ent}), 
such that, $S[i] = \frac{\sum_{(d,c) \in I[i]} \texttt{Ent}(i,d,c,q)}{|I[i]|}$. Here, the function \texttt{Ent} takes the instance $i$, the answer spans $c$ to determine the parent sentence in the text segment $d$, and query $q$ to determine the answer type for the hypothesis.
% Entailment
% . by obtaining the entailment score 
% of the hypothesis, \textit{``(instance) is a (answer type)''}, from the premise, the sentence where the candidate was extracted from, {(entailment higher than contradiction and neutrality)}, indicates type compatibility. We use~\cite{liu2019roberta} to obtain entailment score (\textit{Ent}), such that, $score_i = \sum_{d \in D_i} \textit{Ent}_i^d$.
\end{itemize}
% In the evaluation section, we investigate whether type information and answer confidences add value over simple frequency counts.
%\sg{Remove only context frequency?}
%\squishlist
%\item[5.] \textit{Context Frequency.} We also use the naive approach to extract instances from all $d\in D$ through NER without any QA model, and score each instance by its context frequency, such that, $score_i = \frac{|D_i|}{|D|}$.
%\squishend

% An instance may occur in multiple text segments and we exploit this redundancy using a strength-based ranker similar to~\cite{wang2018evidence}. We score an instance $e$ based on i) its context frequency \emph{i.e.,} the number of text segments containing $e$ in its predicted instances ($|D_e|$), and the confidence of the parent answer span of the instance  $p_e^d$ aggregated over all document occurrences, $d \in D_e$ (see Eq.~\ref{eq:score_e}).
% \begin{equation}\label{eq:score_e}
%     score_e = 0.5\frac{|D_e|}{|D|} + 0.5\sum_{d \in D_e} p_e^d
% \end{equation}

\begin{algorithm}
\caption{Extracting answer explanations}\label{algo:explanation}
% \hspace*{\algorithmicindent} 
\textbf{Input:} Count query, $q$,\\
\hspace*{1cm}set of relevant text segments, $D$,\\
\hspace*{1cm}span predictor model, $\textsc{SpanPrediction}$,\\
\hspace*{1cm}candidate selection threshold, $\theta$,\\
\hspace*{1cm}named-entity recognizer, $\textit{NER}$,\\
\hspace*{1cm}instance consolidation function, $\textsc{ConScore}$\\
% \hspace*{\algorithmicindent} 
\textbf{Output:} A ranked list of instances $I_\textit{Ranked}$
\begin{algorithmic}[1]
    \State $q' \gets q.\textit{replace}(\textit{``how many''}, \textit{``which''})$
    \State $I \gets \{\}$ \Comment{Inverted dictionary of instances.}
    \State $S \gets \{\}$ \Comment{Consolidated score for instances.}
    \For{$d \in D$}
        \State $c \gets \Call{SpanPrediction}{d, q'}$ 
        \LComment{Get all instances from the span.} 
        \If{$c.Span \neq \textit{None and }c.\textit{Conf} > \theta$}
            \State $I_{d} \gets \Call{NER}{c.Span}$ 
            \For{$i \in I_{d}$} 
                \State $I[i] \gets I[i] \cup (d, c)$
            \EndFor
        \EndIf
    \EndFor
    \For{$i \in I$}
        \State $S[i] \gets \Call{ConScore}{I[i], D}$ 
        % \Comment{Return consolidated score for $i$.}
    \EndFor
    \State $I_\textit{Ranked} \gets \Call{SortDescending}{I, \textit{key}=\textit{lambda i}: S[i]}$
    \State \Return $I_\textit{Ranked}$
\end{algorithmic}
\end{algorithm}

\section{The C\lowercase{o}Q\lowercase{u}AD Dataset}\label{sec:dataset}

\subsection{Dataset construction}

\paragraph{Query collection}
Existing QA datasets only incidentally contain count queries; we leverage search engine autocomplete suggestions to automatically compile count queries that reflect real user queries~\citep{how-google-autocomplete-predictions-work}. We provide the Google search engine with iterative query prefixes of the form \textit{``How many \texttt{X}''}, where $X \in \{a, aa, \ldots, zzz\}$, similar to the candidate generation from patterns used in~\cite{romero2019commonsense}, and collect all autocomplete suggestions via SERP API\footnote{\url{https://serpapi.com}}. We keep those with at least one named-entity (to avoid too general queries).
% and no measurement term (to avoid non-entity answer types). This gives us 11.3k count queries. 

\paragraph{Ground Truth Counts} 
We automatically obtain \textit{count ground truth} by collecting structured answers from the same search engine. Executing each query on Google, we scrape knowledge graph (KG) answers and featured snippets, using an off-the-shelf QA extraction model~\citep{sanh2019distilbert} to obtain best answers from the latter. 

We further clean the dataset by removing queries where no counts could be extracted from the text answers and applying simple heuristics to remove queries dealing with measurements. We achieve this by using the CogComp Qualifier which serves a dual purpose. We use it to normalize the text answers to integer counts and identify empty extractions or non-entity answer types when any word representing measurement is returned as \textit{units} of the identified quantity.

This gives us the ground truth for 5k snippet obtained from either KG or from featured snippets. There also exists 4k count queries with no directly available ground truth which we retain in the dataset for evaluation purposes. We manually annotate a sample of 100 queries from those without automated ground truth.

\paragraph{Text Segment Annotation}
We next scrape the top-50 snippets per query from Bing, and obtain 
%\textit{knowledge\_graph} and \textit{answer\_box} in the \textit{json} output, whose values  contain information presented as knowledge panel and featured snippet respectively at the frontend, to categorize the queries based on answer availability. The Google search engine results return knowledge graph answers for 131 queries (CoQuAD$_\text{KG}$), featured snippets for 6.7k queries (CoQuAD$_\text{Featured}$) and vanilla page results with text preview for the remaining 4.5k queries (CoQuAD$_\text{Text}$). We extract the readily available answer span from the knowledge graph (under the keyword \textit{count}) and employ an off-the-shelf QA to extract an answer span from the text of the answer box snippet. We then use a count extractor~\cite{roy2015reasoning} on these answer spans to get a normalized integer count, discarding queries where the answer span contains any measurement units (days, percent, miles, km) as returned by the count extractor or if the quantity is not a positive integer. We do not discard the CoQuAD$_\text{Text}$ since these queries are the ones search engines are worse off and worth evaluating on. 
\textit{text segment ground truth} by labelling answer spans returned by the count extractor~\cite{roy2015reasoning} as positive when the count lies within $\pm 10\%$ from the ground truth. 
There are around 800 queries with no positive snippets, which we do not discard, so the system is not forced to generate an answer. In the end we have $5162$ count queries with automated ground truth, and an average of $40$ annotated text segments per query. 

\paragraph{Evaluation Data}
We use 80\% of the count queries with automated ground truth for training and 20\% for test and development. We report our evaluations on the hand annotated subset of 322 CoQuAD queries which consists both the test data, with the automated ground truths and, queries without any direct answers. We manually annotate the queries with categories, counts, and count contexts. We also track the effect of time, availability of instances and count contexts on these queries. 142 queries that would benefit from instance explanations have at least top-5 prominent manually annotated instances for evaluating answer explanations.

% The test set contains 50 count queries with KG ground truth, 100 with ground truths from featured snippets, and 100 with manually annotated ground truths for quantitative and qualitative analysis. \sg{Include initial CoQuAD 100 query subset}
% Of these 

\subsection{Query Analysis}\label{sec:query_analysis}
\begin{figure}[t]
    \centering
    \includegraphics[width=0.49\textwidth]{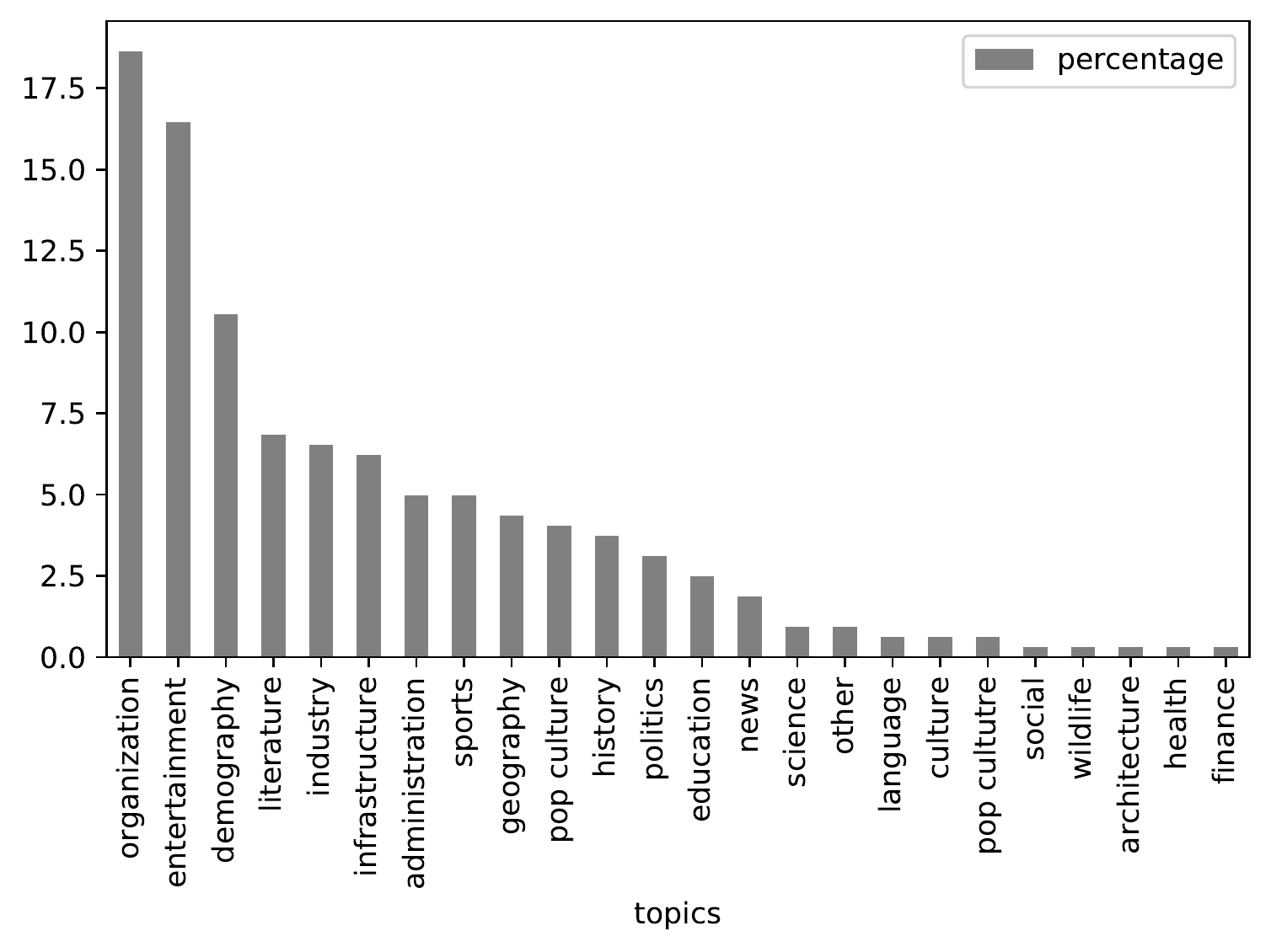}
    \caption{Distribution of topics in CoQuAD. 
    % \sg{Use \%age distribution}
    }
    \label{fig:coquad_topics}
\end{figure}

\begin{figure}[t]
    \centering
    \includegraphics[width=0.49\textwidth]{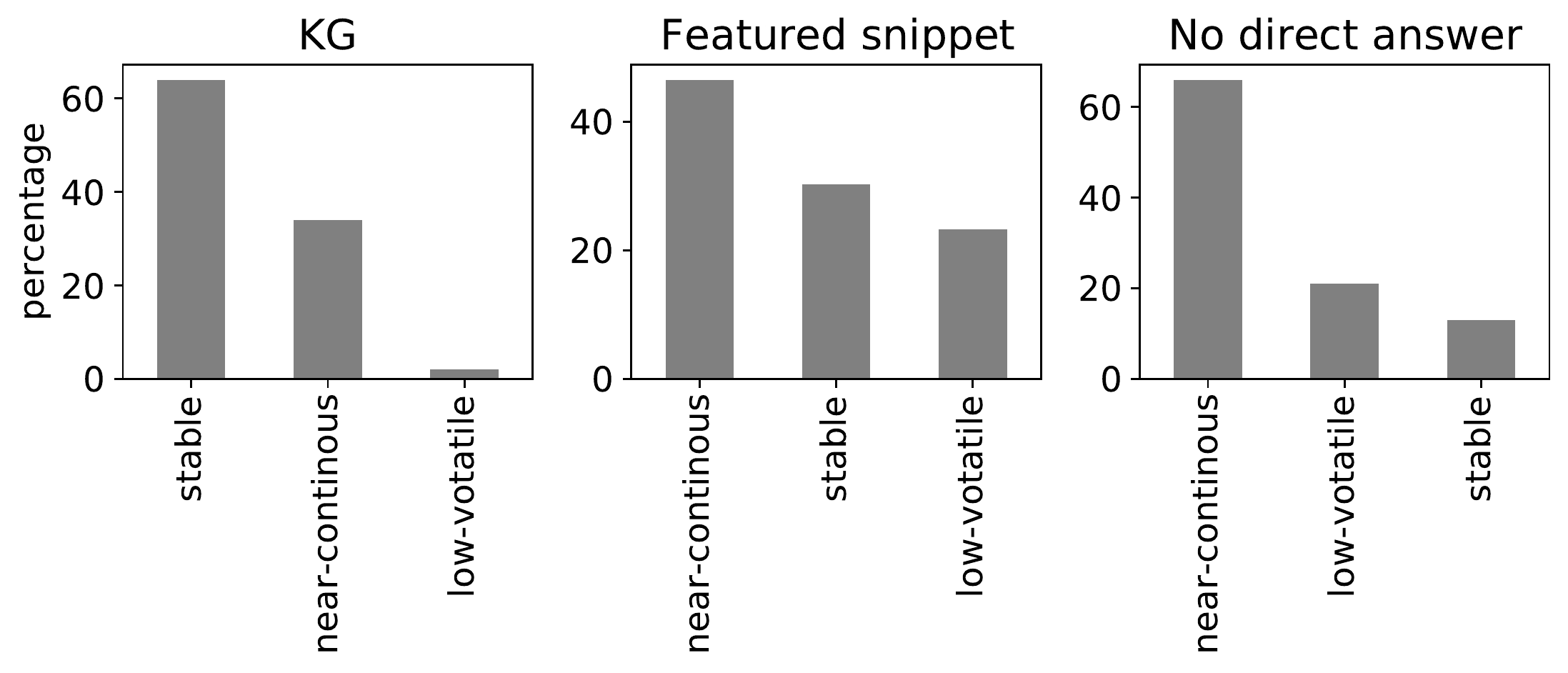}
    \caption{Time variance of CoQuAD queries by answer source. 
    % \sg{Use \%age distribution}
    }
    \label{fig:coquad_topics_timevar}
\end{figure}

Dedicated count question answering is a novel topic for question answering, and as such, we first aim to gain insights into the nature of typical count queries. Our analysis is divided into four questions.
\begin{enumerate}
    \item What ground truths are available for these queries?
    % \item What kind of answers do text snippets provide?
    \item What are the modes of count answers?
    \item What domains do these queries cover, and how topically stable are they?
    \item What are their syntactic characteristics?
\end{enumerate}

We look into the evaluation data of 322 CoQuAD queries to answer these questions unless specified otherwise.

\begin{figure*}
    \centering
    \includegraphics[width=\textwidth,trim={0.5cm 2.7cm 0.3cm 1cm},clip]{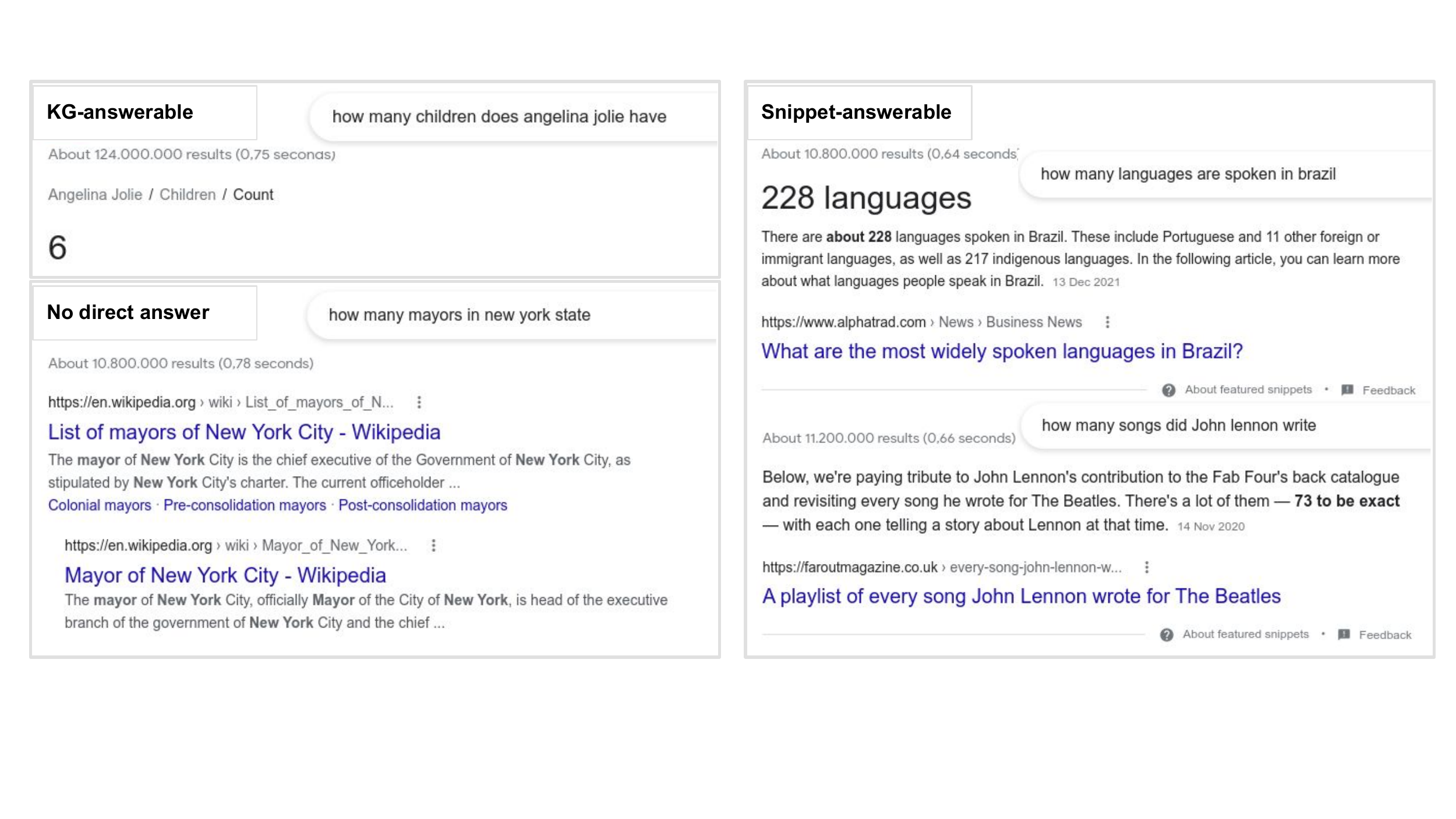}
    \caption{Nature of groundtruths extracted. \textbf{KG-answerable} queries show the path (entity and relation) and aggregate used (Count). \textbf{Snippet-answerable} queries have a featured snippet with a highlighted answer displayed at the top of the snippet (228 languages) or within the snippet which then can be extracted by any off-the-shelf extractive QA models. \textbf{No direct answer} type of queries do not have any automated ground truth as the results returned only ranked.}
    \label{fig:groundtruths}
\end{figure*}

\paragraph{Nature of ground truth}
% KG - 132
% Snippet - 5067
% Organic - 4192

When we automate the ground truth extraction process, we realize that there exists structure to the results provided by the search engines as illustrated in Figure~\ref{fig:groundtruths}. Answers to a small minority of roughly 2\% of the queries come from the internal KG. These \textit{KG-answerable} queries have well-structured outputs. In the case of count queries the answers returned are counts and the path to the KG answer is also displayed to the user. 

The majority of the ground truth labels, (54\% of all CoQuAD queries), are extracted from the top snippet. These snippets rank at the top of the search results, identified as featured snippets and are accompanied by an answer span, highlighted within the text or as a heading of the snippet. We refer to such queries as \textit{snippet-answerable}, and provide two examples in Figure~\ref{fig:groundtruths}.

The queries which yield no automated ground truths (44\% of all CoQuAD queries) come under the \textit{No direct answer} category. As illustrated in Figure~\ref{fig:groundtruths}, these queries only return ranked page snippets, due to the lack of any KG answer or featured snippet.
% \sg{How is this helpful - answers being KG/snippet answerable}

Of the automated ground truth labels, the vast majority come from featured Google snippets, while only 2\% come directly from the Google KG. These labels are thus complemented by a more balanced manual annotation, where we manually labelled 50 questions that were KG answerable, 172 that were snippet-answerable, and 100 without any automated ground truth.

% Around 1\% of the answers came from the Google KG, 59\% through featured snippets, 40\% through page results. 

\paragraph{Answer modes}
% We next analyse the snippets from the search engine, with respect to the modes of answers they provide. 
We have identified three modes how QA systems can answer count queries - via counts, CNPs and instances.

% query_normalization > coquad_characteristics
Since all KG-answerable and snippet-answerable queries have ground truth counts, we focus on the CNPs and instances. 
If we go by the conventional method of QA without any consolidation, and analyse the paths returned by KG answerable queries and the featured snippets, we find that 
KG-answerable queries are usually simply related and have rare occurrences of any semantic qualifiers (equivalent to CNPs) with 0 occurrences in the CoQuAD annotated queries, featured snippets on the other hand contain CNPs in 61\% of the cases. Instances come up in 90\% of the KG-answerable count queries and in only 20\% of the snippet-answerable count queries.

We then proceed by annotating the CoQuAD queries with binary variables indicating whether semantic qualifiers and instance explanations are necessary. For instance, the query \textit{``how many novels did jane austen complete?''} would benefit from instances or CNPs which differentiate her finished and unfinished works or at least hint at the fact. 

We found that indeed helpful semantic qualifiers for KG-answerable queries is necessary in around 8\% of the queries. In snippet-answerable queries and queries with no direct answers, semantic qualifiers are desirable in more than 81.3\% of the cases. As far as instance explanations are concerned, they are desirable in all KG-answerable queries and in 80.8\% of the snippet-answerable queries. We already see a gap between the desired explanatory evidence and what is available when no consolidation is performed.
In Section~\ref{sec:discussion}, we further report on these answer modes in light of the predictions made by {\mysystem} to see whether this gap can be reduced.

% We distinguish four cases:
% \begin{itemize}
%     \item[1.] Instance candidates - 95.3\% of the 322 manually annotated CoQuAD queries have at least one snippet with instance candidates with an average of 15.7 snippets per query containing instance candidates.
%     % (e.g., for ``mayors of New York''),
%     \item[2.] Count candidates - 86.6\% of the 322 queries have snippets with count spans with an average of 5.8 snippets per query containing count candidates.  
%     % (e.g., for ``employees of NHS'') -
%     \item[3.] Both instance and count candidates - 46.8\% of the queries have snippets containing both count and instance candidates with an average of 2.8 snippets per query.
%     \item[4.] Count candidates with semantic qualifiers (e.g., ``7 official languages and 30 regional languagues'') - 86.6\% of the queries have snippets with CNPs, with an average of 5.8 snippets per query containing CNPs. 
%     % #cnps per queries 
% \end{itemize}
 
% We also annotate  
% As discussed earlier \sr{where?},  

% \sr{Numbers needed here!}

%  \textbf{Setting} & \textbf{\#Queries} & \textbf{\#snippets/query} \\
        % Counts for Inference & 258 & 5.24 \\
        % Instance for Explanations & 322 & 35.54 \\
        % Counts \& Instances & 194 & 4.08 \\

\paragraph{Domain and stability}
We assigned high-level topics to the the queries. An expert annotator goes through each of the 322 queries, introducing a new topic label if none of the previous ones make a good match.    
% \sr{a bit more detail here, how many, annotation guidelines}. 
We found that queries in CoQuAD cover a range of topics, notably organizations (18.6\%), entertainment (16.4\%), demography (10.5\%), literature (6.8\%), industry and infrastructure (12.7\%) (see Figure~\ref{fig:coquad_topics}). 

A second important dimension concerns their temporal stability. Queries whose result continuously changes are naturally much harder to deal with, especially if fluctuations are big. We find that 30\% of query results are fully stable (a company's founders, casts in produced movies), 20\% are low-volatile (lakes in a region, band members of an established but active band), 50\% are near-continuous (employment numbers, COVID cases). In Figure~\ref{fig:coquad_topics_timevar} we break it down by answer source (KG, featured snippet, no direct and no direct answers), and we see that the majority of the KG answerable queries is stable (64\%) and near-continuous for the rest. Featured snippet can be used to answer all time-variant queries, though near-continuous queries are a majority (46.5\%). Near-continuous queries form an overwhelming majority (66\%) where search-engines do not provide any direct answers. Only about 13\% of queries with no direct answers are stable, which means that search-engines can handle such queries with little to no-variance.

\paragraph{Syntactic properties}
We identify different query components, namely the named entities, the relation, the type of the entities being counted (conventionally referred to as the type of the answer entity), and remaining context as a bag of words. As discussed in the methodology (Section~\ref{sec:methodology}), we use the types for checking compatibility of the instances as answer explanations. We apply basic natural language processing\footnote{We use SpaCy's \texttt{en\_core\_web\_sm} model.} to get the components. 

\begin{itemize}
\item \textit{Named entity}: was extracted using NER and from tokens with a \texttt{propoer noun} tagged POS label.
\item \textit{Relation}: was identified when a token had a POS tag $=$ \texttt{verb} and the token was the root of the dependency parse tree.
\item \textit{Answer types}: is meant to identify the type of the counted entities by extracting the first \texttt{noun} from the dependency parse tree with any of it's preceding \texttt{adjectives}.
\item \textit{Context}: was all remaining keyword tokens, \emph{i.e.,} excluding conjunctions, determiners, auxiliary verbs, pronouns, punctuation.
\end{itemize}

Most queries count entities in a simple relation to one named entity, i.e., the avg.\ query length is 6.40 words, with an average of 1.08 named entities per query. We found that 95\% of the 322 queries returned non-empty answer types, with more than 200 unique phrases, for instance, \textit{sibling, movie, employee, nfl stadium, real estate agent, czech player}. The queries spanned over 49 different relations.
% (\sg{explain query processing here}). 

% \sr{Something more here? How many nouns/verbs/adjectives? How often temporal keywords, e.g., a year in it? Answer type/relations extracted from queries?}

\noindent Our research data is made publicly available\footnote{\small\url{https://github.com/ghoshs/CoQEx}}.

\section{Experimental Setup}\label{sec:experimental_setup}
% Our evaluation focuses on answering the following four questions.

%\begin{figure}
%    \centering
%    \includegraphics[width=0.4\textwidth]{figures/coquad_topics.pdf}
%    \caption{Distribution of top-10 topics in the evaluation data.}
%    \label{fig:topic_dist}
%\end{figure}
While we can use regular IR metrics of precision and recall for evaluating answer explanations (Mean Average Precision@k, Recall@k, Hit@k and MRR) and accuracy of classification for evaluating count context categories, we need a new metric for counts. This is because counts come with a natural order and distance function (e.g., 507 may be a good answer when the ground truth is 503, but not when it is 234), for which exact String match or embedding distance is not a suitable metric.

\subsection{Evaluation Metrics}

We report the following evaluation metrics for measuring count inference.
\begin{enumerate}
    \item \textit{Relaxed Precision} (RP) is the fraction of answered queries where the prediction lies within $\pm10\%$ of the ground truth and is reported as a percentage.
    \item \textit{Coverage} (Cov) measures the fraction of queries that a systems returns an answer for and is reported as a percentage.
    \item \textit{Relaxed Precision-Coverage trade-off} (P/C) is the harmonic mean of the relaxed precision and coverage values, reported as a percentage.
    \item \textit{Proximity} $\in [0,1]$, which is the ratio of the minimum of the predicted and the gold answer to the maximum of the two, averaged over all queries.
\end{enumerate}

Since we deal with non-canonicalized surface forms of instances, we maintained a list of aliases, obtained from Wikidata, for the annotated prominent instances. 
An instance is relevant if its length-normalized Levenshtein distance~\cite{navarro2001guided} from any of the aliases is less than 0.1.
% We adapted Levenshtein string matching~\cite{navarro2001guided} for an automated evaluation, such that an instance is relevant if its length-normalized Levenshtein distance from any of the aliases less than $0.1$. 
We evaluate answer explanations 
% on queries which benefit from instance explanation and have ground truths
on the following metrics:
% use multiple popular metrics for measuring answer explanations in a more unbiased manner, since different systems prioritize different metrics. To compare performance on answer explanation we use the following metrics.
% \sg{Update with references and justification}
\begin{enumerate}
    \item \textit{Mean Average Precision} (MAP) is the fraction of retrieved entities that are relevant, averaged over the queries.
    \item \textit{Average Recall} (AR) is the fraction of ground truth entities retrieved, averaged over the queries. 
    \item \textit{Hit@k} is the percentage of queries with at least 1 relevant answer in the top-k.
    \item \textit{Mean Reciprocal Rank} (MRR) is the inverse of the rank of the first relevant result averaged across all queries. 
\end{enumerate}

% \sr{Which metric for CNPs?}
The CNPs are evaluated based on the accuracy of the classified labels of \textit{Synonyms}, \textit{Subgroups} and \textit{Incomparables}, measured as the ratio of correct predictions to the total predictions in each class.
% \sg{Report P/R here as well?}

% Readers should note that test count queries include 

\subsection{Baselines}
We compare our proposed system with two complementary paradigms.
\begin{itemize}
    \item[1.] Knowledge-base question answering: QAnswer~\citep{diefenbach2019qanswer}.
    \item[2.] Commercial search engine QA: Google Search Direct Answers (GoogleSDA). In other words, we scrape the structured results from the result page of the Google Search engine.
\end{itemize}

% \noindent\textbf{Q5. How well does {\mysystem} answer count queries?}\newline
For fairness to QAnswer, which specifically deals with count queries by aggregating on top of the SPARQL query, we queried the system\footnote{QAnswer API at \url{https://qanswer-core1.univ-st-etienne.fr/swagger-ui.html}} twice - the original count query (for the count answer) and a modified variant as in Section~\ref{subsec:coex}, i.e., replacing \textit{``how many''} with \textit{``which''}. We then
post-processed the results to extract count and instances.
% Secondly, we post-processed answers, applying additional count aggregation on QAnswer results that were sets, and building scrapers for GoogleSDA answers that came from featured snippets, to increase its recall beyond direct KB answers. %. We only excluded the case of text snippets - considering those would significantly boost GoogleSDA's coverage on Stresstest from 37\% to 60\%, the catch being that the text would require significant post-processing for answer extraction. 
For evaluating instances by GoogleSDA, we post-processed knowledge graph and featured snippet of the search engine result page, keeping items from list-like structures as instances ranked in their order of appearance. %All evaluation is done on 250 manually annotated count queries.
% (50 CoQuAD$_\text{KG}$, 100 CoQuAD$_\text{Featured}$ and 100 CoQuAD$_\text{Text}$). 

\subsection{Datasets}
In order to test the generalizability of {\mysystem} we present the results on count queries from multiple datasets in addition to CoQuAD:
\begin{enumerate}
    \item 100 count queries from an existing dataset LCQuAD \citep{dubey2019lc},
    \item A manually curated dataset of 100 challenging count queries called \textit{Stresstest}, and
    \item 84 count queries found in the Natural Questions \citep{kwiatkowski2019natural} dataset. These queries are similar in nature to our CoQuAD queries (sample of real user queries from Google), but not subject to our own scraping and filtering, and thus provide a corroboration signal for our larger CoQuAD dataset.
\end{enumerate}

\subsection{Implementation Details}
The candidate generation steps for answer inference and explanation uses two instances of a span prediction model~\cite{joshi2020SpanBERT}. The model for answer inference is trained on CoQuAD for 2 epochs, at a learning rate of $3e^{-5}$ using an Adam optimizer. An input datapoint for training consists of a query, a text segment and a text span containing the count answer (empty if no answer). We train over 3 seeds and report the average score on the test data. For getting the instances from the answer spans, we use the pre-trained SpaCy NER model\footnote{\url{https://spacy.io} on the \texttt{en\_core\_web\_sm} model.}. The model for answer explanation is trained on SQuAD 2.0. 

\section{Results}\label{sec:results}

\begin{table*}[t]    
\caption{Comparing baselines on answer inference results (in percentages), where \textbf{RP}$=$ relaxed precision, \textbf{Cov}$=$coverage and \textbf{P/C}$=$relaxed precision-coverage trade-off.
% \sg{update}
}
    \label{tab:different_datasets}
    \begin{adjustbox}{width=0.9\textwidth,center}
    \centering
    \begin{tabular}{|l|lll|lll|lll|lll|}
    \hline
        \multirow{2}{1cm}{\textbf{System}} & 
        \multicolumn{3}{c|}{\textbf{CoQuAD}} & \multicolumn{3}{c|}{\textbf{LCQuAD$_\textit{count}$}} & \multicolumn{3}{c|}{\textbf{Stresstest}} & \multicolumn{3}{c|}{\textbf{NaturalQuestions}}\\
        %  & \textbf{RP} & \textbf{RP} & \textbf{RP}  \\
        % \textbf{Model} & \textbf{CoQuAD$_{100}$} & \textbf{LCQuAD$_\textit{count}$} &\textbf{Handcrafted} \\
        & \textbf{RP} & \textbf{Cov} & \textbf{P/C} & \textbf{RP} & \textbf{Cov} & \textbf{P/C} & \textbf{RP} & \textbf{Cov} & \textbf{P/C} & \textbf{RP} & \textbf{Cov} & \textbf{P/C}\\
        \hline
        QAnswer~\citep{diefenbach2019qanswer} &  6.6 & {96.2} & 12.4 &
        45.0 &  96.1 & \textbf{61.3} & 
        9.0 & 100 & 16.5 & 
        12.5 & {98.8} & 22.1\\
        % GoogleSDA & \textbf{74.7} & 68.9 & 5.7 & 10.5 & 59.0 & 37.0 & \textbf{88.8} & 33.3\\
        GoogleSDA & {93.2} & 18.3 & 30.6 &
        44.4 & 8.6 & 14.4 &
        79.3 & 29.0 & 42.4 &
        {94.4} & 22.6 & 36.4\\
        \hline
        % {\mysystem}~\citep{ghosh2022answering} & 27.4 & {82.0} & 6.7 &  45.2 & {38.6} & {88.3} & 44.2 & 90.0\\
        {\mysystem} & 37.7 & {84.7} & \textbf{52.2} &
        13.6 &  49.3 & 21.3 &
        {43.6} & {91.6} & \textbf{59.1} &
        43.0 & 91.6 & \textbf{58.5}\\
        % CoIn (documents) & \textbf{29.3} & \textbf{60.8} & \textbf{73.9} & 3.2 & 35.7 & 43.2 & 15.3 & 49.4 & 70.6 \\ 
        \hline 
    \end{tabular}
    \end{adjustbox}
\end{table*}

% \begin{table}[t]
%     \caption{Precision@k (P@k), Recall@k (R@k), Hit10 and MRR for the answer explanations of {\mysystem} and baselines.}
%     \label{tab:enum_comparative}
%     \begin{adjustbox}{width=0.49\textwidth,center}
%     \centering
%     \begin{tabular}{|l|l|l|l|}
%     \hline
%     \textbf{Systems} & \makecell{\textbf{QAnswer}\\\citep{diefenbach2019qanswer}}
%     & \textbf{GoogleSDA} & \textbf{\mysystem}\\
%     \hline
%     \multicolumn{4}{|l|}{CoQuAD}\\\hline
%     P@1 & 4.7 & 7.7 & 6.4\\
%     P@5 & 4.5 & 14.2 & 12.8 \\
%     P@10 & 3.7 & 21.7 & 10.8 \\
%     R@1 & 2.1 & 1.3 & 1.3\\
%     R@5 & 0.8 & 2.2 & 11.7\\
%     R@10 & 1.1 & 3.2 & 18.2\\
%     Hit@10 & 12.8 & 12.8 & 55.1\\
%     MRR & 0.064 & 0.089 & 0.218\\
%     \hline
%     \multicolumn{4}{|l|}{NaturalQuestions}\\\hline
%     P@1 & 12.5 & 0.0 & 3.3\\
%     P@5 & 10.0 & 3.3 & 10.0\\
%     P@10 & 7.5 & 0.0 & 9.3\\
%     R@1 & 6.0 & 0.0 & 2.4\\
%     R@5 & 14.3 & 1.2 & 23.8\\
%     R@10 & 14.3 & 1.2 & 29.8\\
%     Hit@10 & 2.0 & 0.0 & 12.6\\
%     MRR & 0.089 & 0.006 & 0.111\\
%     \hline
%     \end{tabular}
%     \end{adjustbox}
% \end{table}

\subsection{Baseline Comparison on Answer Inference}
% For \textit{answer inference}, 
% \paragraph{Answer Inference}
Table~\ref{tab:different_datasets} shows the answer inference performance of {\mysystem} against the baselines on different datasets. We improve upon~\cite{ghosh2022answering} by using a 
% We introduce a larger CoQuAD test set containing 322 queries and the Natural Questions dataset in addition to the previous datasets in~\citet{ghosh2022answering}. 
refined count extractor and also report the performance on a new dataset. The RP-Coverage trade-off metric highlights the advantages provided by {\mysystem}.
% The RP is defined on the number of answered questions, which increases the values w.r.t those in~\citet{ghosh2022answering}.

In both CoQuAD and Natural Questions datasets Google\-SDA has a an RP above 90\%, albeit for very low coverage, whereas QAnswer has a high coverage, more than 96\% in all datasets with poor RP. {\mysystem} not only provides a high coverage, but also a decent RP, with the improved version increasing RP by 10\% on CoQuAD. Except on the LCQuAD dataset, {\mysystem} provides a better trade-off than the baselines.

On the LCQuAD dataset, designed specifically for KG queries, 
% {\mysystem} maintains a coverage close to QAnswer even though it loses in RP. It
 it can be argued that as the LCQuAD queries are created from question templates which in turn are generated from SPARQL templates, the semantic gap between the natural language query and its KG counterpart is much lower. This aids QAnswer and hinders natural language document retrievers used in the other baselines. The fact that {\mysystem} has the lowest coverage in LCQuAD among all datasets also backs this hypothesis.

The manually created Stresstest dataset shows the potential of {\mysystem} in terms of coverage and RP, even though RP of GoogleSDA is higher.
Here results indicate that reliance on structured KBs (QAnswer) is not sufficient for general queries, and robust consolidation
from crisper text segments is necessary.

% While {\mysystem} manages comparable coverage with the baselines, it is far behind in RP in most datasets. In the next part we will look closer into the RP of the baselines by analysing the query type.
% shows that while GoogleSDA has a high precision on CoQuAD (consisting over 200 KG and snippet answerable queries), {\mysystem} not only provides high coverage but a decent RP. On LCQuAD, a dataset designed specifically for KG queries, {\mysystem} loses to LCQuAD while still maintaining a better coverage and RP score compared to GoogleSDA. 

\begin{table*}[t]
    \caption{Comparing answer inference results (in percentages) by GT source of CoQuAD queries: KG-answerable, snippet-answerable and no direct answers (NDA). The number of queries in each type in mentioned in brackets in the column header.
    % \sg{Update}
    }
    \label{tab:inference_query_types}
    \begin{adjustbox}{width=0.7\textwidth,center}
    \centering
    \begin{tabular}{|l|lll|lll|lll|}
    \hline
        \multirow{2}{1cm}{\textbf{System}} &
        \multicolumn{3}{c|}{\textbf{KG} (50)} & 
        \multicolumn{3}{c|}{\textbf{Snippet} (172)} & 
        \multicolumn{3}{c|}{\textbf{NDA} (100)}\\
        %  & \textbf{RP} & \textbf{RP} & \textbf{RP}  \\
        % \textbf{Model} & \textbf{CoQuAD$_{100}$} & \textbf{LCQuAD$_\textit{count}$} &\textbf{Handcrafted} \\
        & \textbf{RP} & \textbf{Cov} & \textbf{P/C} & \textbf{RP} & \textbf{Cov} & \textbf{P/C} & \textbf{RP} & \textbf{Cov} & \textbf{P/C}\\
        \hline
        QAnswer~\citep{diefenbach2019qanswer} & 12.2 & 98.0 & 21.7 &
        4.1 & 97.0 & 8.0 &
        8.2 & 94.0 & 15.1 \\
        GoogleSDA & 100 & 100 & \textbf{100} &
        75.0 & 2.3 & 4.5 &
        40.0 & 5.0 & 8.8 \\
        \hline
        {\mysystem} & 23.1 & 98.0 & 37.4 &
        45.3 & 85.8 & \textbf{59.3} &
        31.9 & 76.3 & \textbf{45.0}\\
        % CoIn (documents) & \textbf{29.3} & \textbf{60.8} & \textbf{73.9} & 3.2 & 35.7 & 43.2 & 15.3 & 49.4 & 70.6 \\ 
        \hline 
    \end{tabular}
    \end{adjustbox}
\end{table*}

\subsection{Effect of Query Types on Answer Inference}
In Table~\ref{tab:inference_query_types}, we analyse how QA systems perform on the answer inference when a query is {KG-answerable},
% \emph{i.e.,} when the result exists in the Google KG, when a query is
{snippet-answerable} 
% \emph{i.e.,} the answer is absent or cannot be extracted directly from a KG but there exists an direct answer snippet 
and when a query is \textit{not directly answerable}. The difficulty in answering the queries increases with each type.
% , \emph{i.e.,} the Google SDA refrained from giving a direct answer.

We observe that the baselines achieve their best performance on KG-answerable queries. While QAnswer has a high coverage, the RP metric is quite low, even for KG-answerable queries. GoogleSDA has by design 100\% precision in KG-answerable queries. While its coverage goes down drastically with increasing difficulty levels of the queries, barely above 5\%, the RP remains respectable. {\mysystem} maintains a decent balance between coverage and RP values in all three scenarios. 

Since, {\mysystem} considers only text, it loses to GoogleSDA in KG-answerable queries by a margin, but is still better than QAnswer. In snippet-answerable queries and queries with no direct answers, {\mysystem} provides a much better P/C trade-off than the baselines.
% \sg{Table~\ref{tab:inference_query_types} goes here.}

% \subsection{Answer Contextualization}
% Table~\ref{tab:different_datasets} is the comparative performance on answer inference. We see that despite high coverage by QAnswer, the relaxed precision is very low even on KG-answerable queries. GoogleSDA has high coverage on CoQuAD$_{KG}$ and CoQuAD$_{Featured}$ but is less precise on KG queries (70\% RP) and miserable on CoQuAD$_{Test}$. Nevertheless, since snippets form the majority of the CoQuAD training data, the high RP of 92\% reinforces the quality of our daaset. {\mysystem} performs decently on coverage, and even if not quite close to GoogleSDA on CoQuAD$_{KG}$ and CoQuAD$_{Featured}$ on RP, it beats QAnswer as the nature of queries become tougher (KG$\rightarrow$Featured Snippet$\rightarrow$Text). 

\subsection{Evaluating Answer Contexualizations}
For evaluating count contextualizations, we cannot directly compare CNPs acquired through {\mysystem} with the other baselines, especially in the KB-QA setting because by nature of the baselines, they return answers with little to no context.

Semantic qualifiers are still common in GoogleSDA featured snippets, coming up in 61\% of queries. While semantic qualifiers can be expressed in KG answers (``volcanic islands in Hawaii''$\Rightarrow$islands$\rightarrow$Hawaii$\rightarrow$volcanoes), this rarely shows up in GoogleSDA for two reasons, i) KG answers are provided for short (single entity) and simple queries (relations with no semantic qualifiers) and ii) KG answers are provided for very popular queried entity and qualifier. 

Unlike the hybrid mode in GoogleSDA which returns results from both a KB and texts, QAnswer is fully KB-based, and SPARQL query understanding is challenging. 
If we only consider the top-1 SPARQL query we get detailed interpretation of the natural language query but the result is homogeneous. For example in the query \textit{``how many territories does canada have''}, \textit{territories} is interpreted as \textit{located in the administrative territorial entity} and in the query \textit{``how many poems did emily dickinson write''}, \textit{write} is interpreted as \textit{author}.

% \subsection{Analyzing Contextualization in KB-QA}
The relations and answer types used in the top-k SPARQL queries can provide insights into existing contextualization in KB-QA as follows. If a subsequent query returns the same set of answers and has similar interpretation, then we have \textit{Synonyms} and when a subsequent query returns an overlapping set of answers such that one query returns subset of the other, we have \textit{Subgroups}. 

When we look into the top-2 SPARQL queries we find that only about 3\% of the queries provide equivalent answers. These however, cannot be considered \textit{Synonyms} by definition since they are always query equivalent reformulations. Typical relations are \textit{spouse} and \textit{sibling} such that the reformulations $\langle?x, spouse, Entity \rangle$ and $\langle Entity, spouse, ?x \rangle$ are symmetric and the answer sets are identical but no additional semantic context is obtained.

In only 5\% of the queries, where one SPARQL query returns the subset of the other, we find distinct relations indicating subgroups. For example \textit{``albums''} in the query, \textit{``how many elton john albums are there''} is interpreted as \textit{album} in the first query and \textit{studio album} in the second and \textit{``mvps''} in \textit{``how many mvps does kobe bryant have''} is interpreted as \textit{NBA Most Valuable Player Award} in the first query and \textit{most valuable player award} in the second. 

\begin{table*}[t]
    \caption{MAP@k, AR@k (R@k), Hit10 and MRR of {\mysystem} and baselines for the answer explanations.
    }
    \label{tab:enum_comparative}
    \begin{adjustbox}{width=0.7\textwidth,center}
    \centering
    \begin{tabular}{|l|llllllll|}
    \hline
        % \multirow{2}{2cm}
        {\textbf{System}} 
        % & \multicolumn{8}{c|}{\textbf{CoQuAD}} & \multicolumn{8}{c|}{\textbf{NaturalQuestions}} \\
        & \textbf{MAP@1} & \textbf{MAP@5} & \textbf{MAP@10}
        & \textbf{AR@1} & \textbf{AR@5} & \textbf{AR@10}
        & \textbf{Hit@10} 
        & \textbf{MRR}
        % & \textbf{P@1} & \textbf{P@5} & \textbf{P@10}
        % & \textbf{R@1} & \textbf{R@5} & \textbf{R@10}
        % & \textbf{Hit10} 
        % & \textbf{MRR}
        \\
        \hline
        \multicolumn{9}{|l|}{CoQuAD (142 queries)}\\
        \hline
        QAnswer~\citep{diefenbach2019qanswer} & 8.5 & 9.3 & 9.6 
        % & 1.7 
        & {2.9} & 6.5 & 8.4
        & 19.7
        & 0.118
        \\
        GoogleSDA & \textbf{14.8} & \textbf{12.8} & \textbf{10.6} 
        % & \textbf{9.5} 
        & \textbf{4.8} & \textbf{13.7} & \textbf{14.3}
        & 23.2
        & 0.185
        \\
        \textbf{\mysystem} & 12.0 & 11.7 & 11.0
        % & {8.3} 
        & 2.3 & 9.3 & 12.7
        & \textbf{37.3}
        & \textbf{0.200} 
        \\ 
        \hline
        \multicolumn{9}{|l|}{Natural Questions (84 queries)}\\
        \hline 
        QAnswer~\citep{diefenbach2019qanswer}
        & 14.3 & 15.7 & 16.3 
        & 5.0 & 11.1 & 14.3 
        & 33.3
        & 0.199 \\
        GoogleSDA
        & \textbf{25.0} & \textbf{21.7} & \textbf{17.9} 
        & \textbf{8.0} & \textbf{23.2} & \textbf{24.2} 
        & \textbf{39.3} & \textbf{0.313} \\
        \textbf{\mysystem} 
        & 9.5 & 8.0 & 7.2 
        & 3.6 & 8.0 & 11.2 
        & 25.0 & 0.143
        \\
        \hline
    \end{tabular}
    \end{adjustbox}
\end{table*}

\subsection{Baseline Comparison on Answer Explanation}
% \subsection{Answer Explanations}
% For \textit{answer explanation},
The results on instance-annotated CoQuAD and natural Questions dataset are in Table~\ref{tab:enum_comparative}. We see that GoogleSDA is the best across datasets in terms of MAP and AR. {\mysystem} comes close in the CoQuAD dataset but performs worse than both baselines in the Natural Questions dataset. Instance explanations when readily available in KGs can be extracted with a single query. Texts prove useful when KG is incomplete or the SPARQL translation is not capture the user intent. We test this hypothesis in the next sub-section.

% {\mysystem} provides more R than the baselines, with competitive P at rank 1, and better at ranks 5 and 10. {\mysystem} performs consistently better at hits@k and MRR, losing only slightly in precision at ranks 1 and 5. Both baselines answer less than 25 queries at rank 5, and at rank 10 less than 20 queries, and QAnswer performs poorly in the returned answers. GoogleSDA operates extremely conservative, thus maintaining precision at lower ranks, at the cost of tiny recall.

% Compared to the answer inference task, answer explanation scores are quite low. The low recall scores can be explained by the incomplete ground truth annotations. 
We identify some challenges which can be tackled to improve instance explanations from text.
The low precision scores of {\mysystem} can be attributed to i) noise due to non-entity terms recognized as entities ii) alternate human-readable surface forms like the \textit{``European Union''} as the group with which \textit{``South Korea''} has a foreign trade agreement with instead of the specific group name \textit{``European Free Trade Association''}, iii) entities satisfying a more general criteria, for instance returning other airports from Vietnam when asked for airports in Ho Chi Minh City, iii) local or generalized surface forms, for instance \textit{``Himalayan rivers''} referring to the group of rivers in India originating from the Himalayan mountain range instead of specifically naming the rivers. These errors are specific to texts and are difficult to overcome without human annotation. 

% A case of false negatives in {\mysystem} arises in cases when the entities returned are missing from gold annotation. For instance, {\mysystem} returns \textit{mauna kea} and \textit{diamond head} (absent from the annotated corpus) in addition to \textit{haleakala} (in the annotated corpus), all dormant or extinct volcanoes for the query on \textit{inactive volcanoes in Hawaii}. The missing entities are found to be correct when a reverse search is done on the entities but are not directly available to the user through top search results. False negatives occur in the baselines as well, but they are less interesting because, this can be rectified by adding more gold annotations, whereas with {\mysystem} there are cases where instances buried in text are easily available to the user.    

\begin{table*}[t]
    \caption{MAP@k, AR@k (R@k), Hit@10 and MRR for the answer explanations of {\mysystem} and baselines on CoQuAD queries by their GT source.
    }
    \label{tab:enum_query_type}
    \begin{adjustbox}{width=0.7\textwidth,center}
    \centering
    \begin{tabular}{|l|llllllll|}
    \hline
        % \multirow{2}{2cm}
        {\textbf{System}} 
        % & \multicolumn{8}{c|}{\textbf{CoQuAD}} & \multicolumn{8}{c|}{\textbf{NaturalQuestions}} \\
        & \textbf{MAP@1} & \textbf{MAP@5} & \textbf{MAP@10}
        & \textbf{AR@1} & \textbf{AR@5} & \textbf{AR@10}
        & \textbf{Hit@10} 
        & \textbf{MRR}
        % & \textbf{P@1} & \textbf{P@5} & \textbf{P@10}
        % & \textbf{R@1} & \textbf{R@5} & \textbf{R@10}
        % & \textbf{Hit10} 
        % & \textbf{MRR}
        \\
        \hline
        \multicolumn{9}{|l|}{KG (50 queries)}\\
        \hline
        QAnswer~\citep{diefenbach2019qanswer} 
        & 20.0 & 21.3 & 22.0
        & 6.1 & 14.2 & 18.5
        & 38.0
        & 0.250
        \\
        GoogleSDA 
        & \textbf{42.0} & \textbf{36.4} & \textbf{30.0}
        & \textbf{13.5} & \textbf{38.9} & \textbf{40.7}
        & \textbf{66.0}
        & \textbf{0.526}
        \\
        \textbf{\mysystem}
        & 14.0 & 13.7 & 12.9
        & 3.8 & 13.0 & 18.4
        & 42.0
        & 0.233
        \\ 
        \hline
        \multicolumn{9}{|l|}{Snippet (92 queries)}\\
        \hline 
        QAnswer~\citep{diefenbach2019qanswer}
        & 2.2 & 2.8 & 2.9
        & 1.2 & 2.4 & 3.0 
        & 9.8 
        & 0.046\\
        GoogleSDA
        & 0.0 & 0.0 & 0.0 
        & 0.0 & 0.0 & 0.0
        & 0.0 
        & 0.0\\
        \textbf{\mysystem} 
        & \textbf{10.9} & \textbf{10.6} & \textbf{10.0}
        & \textbf{1.5} & \textbf{7.3} & \textbf{9.7}
        & \textbf{34.8}
        & \textbf{0.182}
        \\
        \hline
        % \multicolumn{9}{|l|}{No direct answer}\\
        % \hline 
        % QAnswer~\citep{diefenbach2019qanswer}
        % & &&&&&&&\\
        % GoogleSDA
        % & &&&&&&&\\
        % \textbf{\mysystem} 
        % & &&&&&&&
        % \\
        % \hline
    \end{tabular}
    \end{adjustbox}
\end{table*}

\subsection{Answer Explanation by Query Type}
% As in Table~\ref{tab:inference_query_types}, 
We analyse the system performances on answer explanation by the query answerability: {KG-answerable}, {snippet-answerable} in Table~\ref{tab:enum_query_type}. Our hypothesis is that the baselines perform very well on answer explanations when the queries are KG-answerable. The performance values support this since we observe that Google\-SDA provides the best precision, recall and MRR sores, followed by QAnswer for the KG-answerable queries. Given that CoQEx only uses text information, it still finds relevant instances achieving a 14\% MAP at rank 1.

In the case of snippet-answerable queries, the dependence of the baselines on KGs becomes clear. CoQEx performs the best followed by QAnswer and GoogleSDA. It should be noted that GoogleSDA might return list pages in the search result, such that if we were to scrape the contents of the list page, we would likely find correct instances. However, 
% these are results that Google has lower confidence in (shown by the emphasis put on not highlighting a direct answer in these tables), thus 
we limit ourselves to instances found on the result page itself, either as an answer from its KG or in the form of direct answers (featured snippets).  

% \sg{Table~\ref{tab:enum_query_type} goes here.}

\begin{table}[t]
    \caption{User preference for different explanation modes (in percentages).}
    \label{tab:user_preference}
    \begin{adjustbox}{width=0.49\textwidth,center}
    \centering
    \begin{tabular}{|l|llll|}
    \hline
        \textbf{Explanation Type} & \textbf{Bare Count} & \textbf{Explanation} & \textbf{Both} & \textbf{None} \\
        \hline
        {CNPs} & 13.3 & 50.0 & 33.3 & 3.4 \\
        {Instances} & 3.3 & 73.3 & 23.4 & 0.0 \\
        {Snippet} & 0.0 & 80.0 & 20.0 & 0.0 \\
        {All} & 10.0 & 63.3 & 23.4 & 3.3 \\
        \hline
    \end{tabular}
    \end{adjustbox}
\end{table}

\begin{table}[t]
    \caption{Extrinsic user study on annotator precision (in percentage).}
    \label{tab:user_study_eval}
    \begin{adjustbox}{width=0.49\textwidth,center}
    \centering
    \begin{tabular}{|l|lllll|}
    \hline
    \textbf{Class} & \textbf{Only Count} & \textbf{+Instances} & \textbf{+CNPs} & \textbf{+Snippet} & \textbf{All} \\
    \hline
    Correct & 73 & 63 & 78 & 75 & \textbf{88}\\
    Incorrect & 28 & 45 & 40 & \textbf{53} & 45\\
    Both & 55 & 56 & 63 & 66 & \textbf{71} \\
   \hline
    \end{tabular}
    \end{adjustbox}
\end{table}

\subsection{User Studies}
To further verify the user perception of our enhanced answers, and their extrinsic utility, we performed two user studies.

\paragraph{User study 1: Intrinsic answer assessment}
We asked 120 MTurk users for pairwise preferences between answer pages that reported bare counts, and counts enhanced by either of the explanation types. The preference of different explanation types are shown in Table~\ref{tab:user_preference}. 50\% of participants preferred interfaces with CNPs, 80\% with a snippet, 73\% with instances, 63\% preferred an interface with all three enabled.
While snippets are already in use in search engines, the results indicate that CNPs and instances are considered valuable, too. 

% \sg{Compare on all 10 queries?}

\paragraph{User study 2: Extrinsic utility for assessing system answer correctness}
We also validated the merit of explanations extrinsically. We took 5 queries with correct count results, 5 with incorrect results, and presented the system output under the 5 explanation settings to 500 users. The users' task was to judge the count as correct or not based on the explanations present. The measured precision scores are in Table~\ref{tab:user_study_eval}. All explanation had a positive effect on overall annotator precision, especially for incorrect counts.

\section{Component analysis of {\mysystem}} 
We evaluate the {\mysystem} components to determine the best configurations for answer inference, consolidation and explanation.  
% We see the importance on \textit{Proximity} subsequently.
% \sg{Report coverage for answer explanations too.}

\begin{figure}[t]
    \centering
    \includegraphics[width=0.49\textwidth]{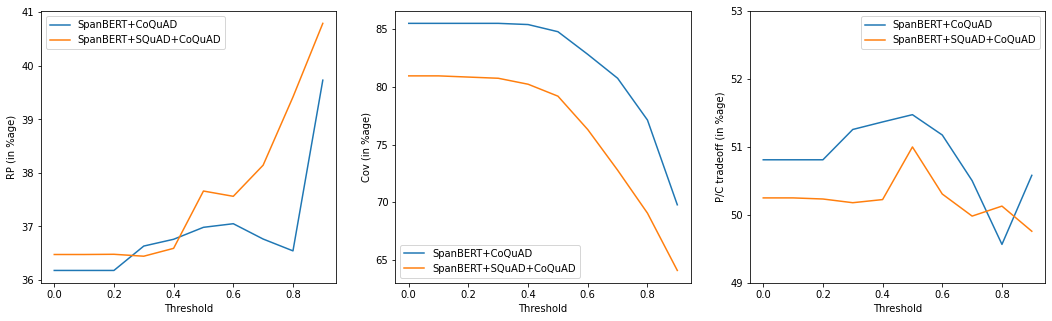}%
    \caption{Performance of fine-tuned models on answer inference metrics, Relaxed Precision (RP), Coverage (Cov) and Relaxed Precision-Coverage trade-off (P/C) across different span selection thresholds.}
    \label{fig:answer_inference_by_model_threshold}
\end{figure}

\subsection{Span Prediction Model for Answer Inference}
We test the candidate generator for count spans on SpanBERT~\citep{joshi2020SpanBERT} finetuned on
% i) the vanilla BERT model~\cite{joshi2020SpanBERT}, 
i)  CoQuAD and,
ii) the popular general QA dataset SQuAD~\citep{rajpurkar2016squad} on different span selection thresholds in Figure~\ref{fig:answer_inference_by_model_threshold}. Span selection works such that counts coming from spans with a model confidence above the threshold is used for aggregation. 
As is expected, the precision goes up while the coverage decreases as the thresholds are set higher, since the model becomes more conservative on high confidence predictions. A threshold of 0.5 gives the best precision-coverage trade-off.

% and iii) on two text inputs i) snippets ii) full web documents. 
% Fine-tuning with CoQuAD performs very close (17.4\% RP) to the SQuAD fine-tuned model (18.9\% RP) on text inputs, even exceeding by 6.5\% to 22.7\% RP on full web documents, which implies that given more context, {\mysystem} is capable of giving better results.
\textit{Fine-tuning} on SQuAD gives higher precision scores at thresholds greater than 0.4. However, this difference, which goes up to 3\% max, is a trade-off to the higher coverage of CoQuAD gives, between 5\%-8\% higher, resulting in overall more correctly answered queries. Here, we average the metrics over all consolidation strategies (\textit{Most Frequent}. \textit{Median} and \textit{Weighted Median}) and compare the consolidation schemes next. 

% \sg{Include BERT base?}

\begin{table}[t]
    \caption{Intrinsic evaluation of the Answer Inference on consolidation alternatives. The model is SpanBERT+CoQuAD with span prediction threshold$=$0.5}
    \label{tab:component_eval_answer_inference}
    \begin{adjustbox}{width=0.49\textwidth,center}
    \centering
    \begin{tabular}{|l|lll|}
        \hline
        \textbf{Consolidation} & \textbf{Relaxed Precision (\%)} & \textbf{Coverage (\%)} & \textbf{Proximity}\\
        % \hline
        % \multicolumn{4}{|l|}{Model Finetuning} \\
        % \hline
        % SpanBERT+SQuAD & 36.0 & 74.7 & 0.606\\ 
        % SpanBERT+CoQuAD & 33.8 & 82.0 & 0.598 \\
        % \hline
        % \multicolumn{4}{|l|}{Consolidation} \\
        \hline
        Median & 35.4 & 84.7 & 0.611\\
        Most Confident & 37.0 & 84.7 & 0.600\\
        Most Frequent & 37.7 & 84.7 & 0.611\\
        \textbf{Weighted Median} & 37.7 & 84.7 & 0.620\\
        \hline
    \end{tabular}
    \end{adjustbox}
\end{table}

\subsection{Best Consolidation for Answer Inference}
When selecting a consolidation strategy, we compare the \textit{Relaxed Precision} and \textit{Proximity} metrics, since coverage is same for all strategies (see Table~\ref{tab:component_eval_answer_inference}).
% For \textit{answer inference}, we see that while the \textit{Confident} consolidation scheme is ahead in RP, it provides very similar \textit{Proximity} values for much lower coverage, 19\% points behind the other consolidation schemes (see Table~\ref{tab:component_eval_answer_inference}).
The \textit{confident} strategy, which in essence performs no consolidation has lower \textit{Proximity} than all consolidation strategies, beating only the \textit{median} consolidation strategy in \textit{Relaxed precision}, where the \textit{confident strategy} leads by 1.6\%.

The \textit{weighted median} is the winning indicating that using model confidence as weights boosts performance. The naive \textit{frequent} strategy comes very close to the \textit{weighted median} scheme, both in terms of RP, where it is equal, and \textit{Proximity} (behind by 0.09). 
% The \textit{frequent} and \textit{weighted median} consolidation schemes outperform the others, with \textit{weighted median} achieving 27.4\% RP just 0.6\% ahead of \textit{frequent}. 
Thus, for queries backed by less variant data, \textit{frequent} is good enough, but to have an edge in more variant data \textit{weighted median} is the way to go.

\subsection{Accuracy of Answer Contextualization}
The accuracy of the CNP categories is directly dependent on the quality of the prediction. Since in this experiment we only want to test the accuracy of the CNP category classifier, we restrict ourselves to CNPs from correct predictions (RP=1). We assess the classification accuracy of CNPs for a manually labelled sample of 601 CNPs for 106 queries. 

While a strict synonym threshold of $\alpha=0\%$ (CNPs equal to predicted count with cosine similarity $>0$),
% \sr{threshold on what? reader certainly has forgotten here} 
ensures a high accuracy of 82.1\% for \textit{Synonyms}, which only decreases with increasing $\alpha$ down to 69.1\% for $\alpha=100\%$, the accuracy of \textit{Subgroups} is initially low (34.4\%), but increases with higher levels of $\alpha$, peaking at $\alpha=60\%$ and then decreases. This happens because as $\alpha$ increases, more incorrect CNPs are classified as \textit{Synonyms}. 

While the \textit{Subgroups} may benefit at lower $\alpha$ values when CNPs with counts a bit further away from the predicted count are correctly classified as \textit{Synonyms}, at higher $\alpha$ the CNPs with very low counts are still mis-classified as \textit{Subgroups} if they have a high semantic similarity to the representative CNP.

The number of \textit{Incomparable} CNPs decreases with increasing $\alpha$, which gives a higher accuracy but at the cost of incorrect \textit{Synonyms} and \textit{Subgroups}.
% while the CNPs in the other classes decrease
% thereby increasing the chances of a higher accuracy. 
A weighted optimum is reached at $\alpha=30\%$, where the accuracy of the \textit{Synonyms} does not degrade much (79.6\%), and the accuracy of \textit{Subgroups} and \textit{Incomparables} is both above 60\% (61.9\% and 71.4\% respectively).

\begin{figure}[t]
    \centering
    \includegraphics[width=0.49\textwidth]{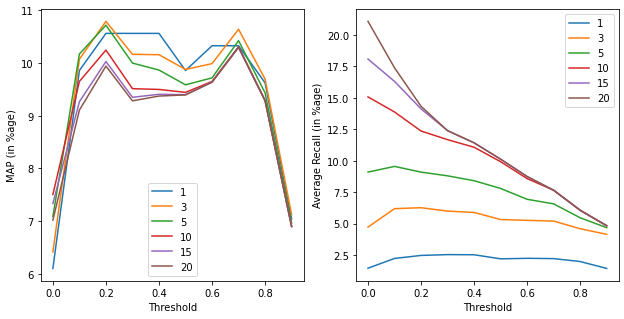}
    \caption{MAP and Average recall across threshold values and ranks. The values are averaged over consolidation alternatives.}
    \label{fig:answer_explantion_by_threshold}
\end{figure}
\subsection{Effect of Model Confidence on Answer Explanation }

% \sr{Absolute precision scores are pretty low 10 percent means 9 out of 10 instances are wrong - what is going on? This should be discussed somewhere, preferably a section that we claim as new material (so not in 7.4) - maybe here or discussion? or in 7.4}

We see the effect on MAP and average recall when we threshold the answer spans from the span prediction model by the model confidence in Figure~\ref{fig:answer_explantion_by_threshold}. We observe that, while the recall goes down in general with increasing model confidence, except for recall at rank 1 which is more or less constant for thresholds between 0.1 and 0.8, the precision drops sharply when model is both less and more confident and has two peaks at 0.2 and 0.7. 

We can argue that the gradual drop in recall is because the model predicts less number of high confidence spans. The precision has a sharp increase initially when increasing the model confidence threshold from 0 to 0.2, because at threshold$=$0, we consider all predictions and this introduces a lot of noise. Whereas, the drop in precision at thresholds 0.8 and higher is probably due to the model being penalized for being too conservative and making sparse or no predictions. Choosing a model confidence threshold of $0.2$, creates a balance between precision and recall values.

% \begin{table}[t]
%     \caption{Intrinsic evaluation of Answer Explanation component across consolidation strategies on Hit@k and Mean Reciprocal Rank (MRR).}
%     \label{tab:component_eval_answer_explanation}
%     \begin{adjustbox}{width=0.3\textwidth,center}
%         \centering
%         \begin{tabular}{|l|ll|}
%             \hline
%             \textbf{Consolidation Strategy} & 
%             % \textbf{P@1} & \textbf{P@5} & \textbf{P@10} &
%             % \textbf{R@1} & \textbf{R@5} & \textbf{R@10} &
%             \textbf{Hit@10} & \textbf{MRR} \\
%             \hline
%             QA w/o Consolidation & 
%             % 11.9 & 4.8 & 2.3 &
%             % 11.8 & 33.8 & 46.3 &
%             5.4 & 0.226 \\
%             QA + Context Frequency &
%             % 7.5 & 10.9 & 9.2 &
%             % 7.4 & 38.2 & 48.5 &
%             17.6 & 0.198\\
%             QA + Summed Confidence &
%             % 5.2 & 8.1 & 7.7 &
%             % 5.1 & 25.7 & 40.4 &
%             14.0 & 0.144\\
%             QA + Type Compatibility &
%             % 6.7 & 8.1 & 7.4 &
%             % 6.6 & 30.9 & 44.9 &
%             14.7 & 0.169\\
%             \hline
%         \end{tabular}
%     \end{adjustbox}
% \end{table} 

\begin{figure}[t]
    \centering
    \includegraphics[width=0.49\textwidth]{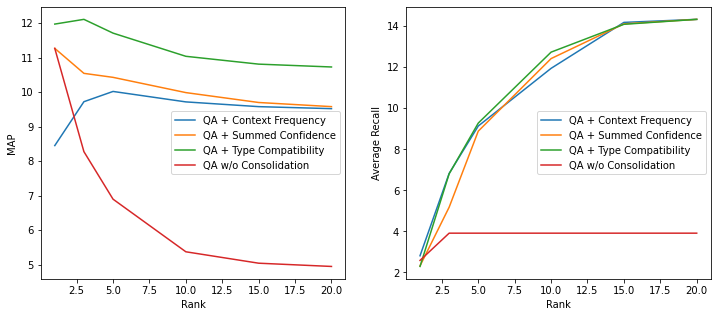}
    \caption{MAP and Average Recall of different consolidation strategies at ranks 1, 5 and 10 when span selection threshold=0.2.}
    \label{fig:answer_explanation_by_consolidation}
\end{figure}

\subsection{Best Consolidation for Answer Explanations}
% \sg{Update: ablations on i) threshold selection (todo), ii) aggregation (complete)}
% In \textit{answer explanation} in Table~\ref{tab:component_eval_answer_explanation}, \textit{QA + context frequency} consolidation performs consistently well. While \textit{QA w/o consolidation} has highest P (11.9\%) and R (11.8\%) at rank 1, its performance decreases subsequently to only 2.3\% P
% % , while maintaining a good recall of 46.3\% R 
% at rank 10, compared to 9.2\% P 
% % and 18.2\% R 
% of \textit{QA + context frequency}. The latter also maintains the highest recall in subsequent ranks. This indicates that the QA model is tailored to the typical setting QA of a single correct answer, and that consolidation helps beyond that. Additionally, consolidation on the instance frequency performs surprisingly better than more sophisticated methods like type compatibility and answer confidence. One reason may be that textual entailment models may need more training for such fine-grained type entailment. Also, depending on model confidences may be misleading when models give high confidence incorrect answers.
% \sg{rewording needed here}
The MAP and AR scores of different consolidation strategies are shown in Figure~\ref{fig:answer_explanation_by_consolidation}. Without consolidation, MAP and average recall are comparable only at rank 1, after which the gap between strategies with and without consolidation increases sharply. All consolidation strategies perform similarly in average recall and the discriminating factor is in the MAP by rank. \textit{QA + Typed Compatibility} is the best overall followed by \textit{QA + Summed Confidence}. The naive \textit{QA + context frequency} performs worse at higher ranks implying that the most frequent named-entities may not be the correct explanations.

Thus we can say that consolidation is a determining factor in increasing performance, with different starting points at rank 1, but converging at higher ranks. \textit{QA + Type Compatibility} is the most stable across ranks, followed by other consolidation strategies. With no consolidation, MAP decreases rapidly across ranks, unlike consolidation methods where MAP is stable across ranks. The recall without any consolidation very quickly stagnates at quite low values (3.3\%). 
% \sg{Precision-Recall scores for different (higher) thresholds for selecting instances from QA models}

% \sg{Introduce tables with performance of all settings/cases in the three components}

\section{Discussion}\label{sec:discussion}

\begin{table}[t]
    \caption{Number of queries with at least 1 contributing snippet under different settings and number of contexts per query satisfying the setting.}
    \label{tab:coexistence_counts_ents}
    \begin{adjustbox}{width=0.49\textwidth,center}
    \centering
    \begin{tabular}{|l|ll|}
        \hline
        \textbf{Setting} & \textbf{\#Queries} & \textbf{\#snippets/query} \\
        \hline
        % \multicolumn{3}{|c|}{Move these to Sec 5.2}\\
        Counts Candidates & 279 & 5.8 \\
        Correct Counts & 106 & 6.0 \\
        Relevant CNPs & 106 & 6.0\\
        Instance Candidates & 307 & 15.7 \\
        Relevant Instances & 39 & 3.9\\
        Counts \& Instance Candidates & 151 & 2.8 \\
        Correct Counts + Relevant Instances & 3 & 1\\
        \hline
    \end{tabular}
    \end{adjustbox}
\end{table}

\subsection{Distribution of CNPs and instances in snippets}
In Section~\ref{sec:query_analysis} we introduced the different answer modes in count queries. An open question is how often these are actually present in text sources.
% \sg{rewrite this and link to Sec 5.2: Answer Types}
We distinguish four notable cases {\mysystem} encounters in the snippets:
\begin{itemize}
    \item[1.] Instance candidates - 95.3\% of the 322 manually annotated CoQuAD queries have at least one snippet with instance candidates with an average of 15.7 snippets per query containing instance candidates.
    % (e.g., for ``mayors of New York''),
    \item[2.] Count candidates - 86.6\% of the 322 queries have snippets with count spans with an average of 5.8 snippets per query containing count candidates.  
    % (e.g., for ``employees of NHS'') -
    \item[3.] Both instance and count candidates - 46.8\% of the queries have snippets containing both count and instance candidates with an average of 2.8 snippets per query.
    \item[4.] Count candidates with semantic qualifiers (e.g., ``7 official languages and 30 regional languagues'') - 86.6\% of the queries have snippets with CNPs, with an average of 5.8 snippets per query containing CNPs. 
    % #cnps per queries 
\end{itemize}

Now that we have established that snippets are a good source of candidates, we also report on the number of queries and number of snippets per query which contain counts that lead to correct predictions and relevant instances. These are summarized in Table~\ref{tab:coexistence_counts_ents}. 
% As discussed in Section ~\ref{sec:query_analysis}, if we only focus on answers without any consolidation, supporting evidence in the form of CNPs are sparsely present, even though CNPs are desired in most queries. {\mysystem} has a high answer coverage returning answers for 84.7\% of the CoQuAD queries. For the unanswered queries, no CNPs are returned since there is no prediction.  
% {\mysystem} predicts an average of 6 CNPs per query for the queries it answers (see Table~\ref{tab:coexistence_counts_ents}). As we see from the user studies CNPs not only enhance correct predictions but they also allow users to spot incorrect predictions.
% \subsection{Distribution of instances in snippets}
% On analysing the KG and snippet answerable CoQuAD questions and their answers, we found that 90\% of the KG answerable queries returned instances and for the snippet answerable queries, 20\% of the featured snippets contained instances. In both cases, instances as means of explanations is required in 100\% and 80\% of the cases, respectively. As we have already stated if the information is in the KG, extracting the information is relatively straightforward, whereas extracting instances from texts is non-trivial. 
Using {\mysystem} we are able to identify relevant instances for 39 of our queries spread across an average of 3.9 snippets per query. Relevant CNPs and counts could be identified in more than 30\% of the count queries with counts spread across an average of 6 snippets per query. 
% Nevertheless, our precision (Table~\ref{tab:enum_comparative}) could be further improved.

\subsection{Coexistence of counts and instances}
In {\mysystem} the tasks concerning counts and instances are separately tackled and a natural question arises as to how counts and instances are spread across the snippets and whether they frequently coexist in the same document, like (``\textit{He wrote 73 songs, for example Let it Be, ...'')}. Frequent coexistence would be very beneficial for the approach, since it would allow focusing on identifying snippets that solve both sub-problems at once.

If we count the queries which contain at least on1 snippet with both counts and instance candidates we find that around 46.8\% (151) queries satisfy this and even the number of queries where correct counts and relevant instances co-occur is only 3 (see table~\ref{tab:coexistence_counts_ents}). The number of snippets per query containing co-occurring counts and instances is also less than 3.
% We look into our CoQuAD dataset of 322 queries and summarize our findings in Table~\ref{tab:coexistence_counts_ents}. 
% We see that separately more than 80\% (258) of the queries have 5 snippets per query on an average, contributing to the counts  which are subsequently used for Answer Inference. All queries have 35 snippets on an average with instances ranked for Answer Explanations (\sg{recheck numbers: any changes after explanation threshold?}). However the number of queries with snippets containing both count and instances is sparse
This indicating that relevant information is spread across contexts making our task of inferring answer and explanatory evidence from multiple sources a significant contribution.   
% is much lower and this further reduces to just 2 queries which have correct count inference and relevant instance such that both come from a common text segment, and this happens in only one text segment.
% \sg{How many snippets have both counts and instances, can one be used to boost the other? Why? Why not?}

\newcounter{rowno}
\setcounter{rowno}{0}
\begin{table*}[th]
    \caption{Example outputs of {\mysystem} with confidence scores of CNPs and aggregated scores of instances in subscripts.}
    % \caption{Count inferences by CoIn and count explanations by CoEx on CoQuAD and Stresstest queries.}
    \label{tab:full_example}
    \centering
    \begin{adjustbox}{width=0.9\textwidth,center}
    \begin{tabular}{|>{\stepcounter{rowno}\therowno.}lp{4.2cm}|l|p{7cm}|p{4cm}|}
    % \begin{tabular}{|p{4cm}|l|p{10cm}|p{3.7cm}|}
        \hline
        \multicolumn{1}{|l}{\textbf{No.}} & 
        \textbf{Query} & \textbf{Inference} & 
        {\textbf{CNPs}} &
        {\textbf{Top-5 Instances}}\\
        \hline
        
        & how many songs did john lennon write for the beatles & 73 & 
        \textbf{CNP$_\textit{rep}$:} 73 songs\subscriptscore{0.92} \newline 
        \textbf{Synonyms:} 61 songs\subscriptscore{0.77} \newline
        \textbf{Subgroups:} 22 songs\subscriptscore{0.67} \newline
        \textbf{Incomparables:} 189 songs\subscriptscore{0.82}, 229 original songs\subscriptscore{0.55}, 229 songs\subscriptscore{0.5} & 
        \textbf{x} John Lennon's\subscriptscore{0.71}  \newline 
        \textbf{x} Beatles\subscriptscore{0.55}  \newline
        \cmark Maggie Mae\subscriptscore{0.54}
        % A Hard Day’s Night\subscriptscore{0.01}
        \\\hline
        
        & how many main islands in hawaii
        % \newline No direct answer \newline answer type has qualifier 
        & 8 &
        \textbf{CNP$_\textit{rep}$:} eight principal islands\subscriptscore{0.96}\newline  
        \textbf{Synonyms:} eight main islands\subscriptscore{0.91}, six major islands\subscriptscore{0.91}, 8 main islands\subscriptscore{0.9}, 8 largest\subscriptscore{0.83} &
        \cmark the Big Island\subscriptscore{0}
        \\\hline 
        
        & how many languages are spoken in indonesia 
        % \newline high variance answer, live query 
        & 709 &
        \textbf{CNP$_\textit{rep}$:} 709 living languages\subscriptscore{0.97} \newline
        \textbf{Synonyms:} 653 languages\subscriptscore{0.98}, estimated 700 languages\subscriptscore{0.91}, 700 living languages\subscriptscore{0.96}, 725 languages\subscriptscore{0.78}, 800 languages\subscriptscore{0.61} \newline
        \textbf{Subgroups:} 300 different native languages\subscriptscore{0.94} &
        \cmark Malay-Indonesian\subscriptscore{0.79} \newline
        \cmark Indonesian language\subscriptscore{0.77} \newline
        \cmark Bahasa\subscriptscore{0.7} \newline
        \cmark Indonesian\subscriptscore{0.35}
        
        \\\hline
        
        & how many osmond brothers are still alive 
        % \newline instances are more helpful
        & 9 &
        \textbf{CNP$_\textit{rep}$:} nine Osmond siblings\subscriptscore{0.89} \newline
        \textbf{Synonyms:} nine siblings\subscriptscore{0.87}, nine children\subscriptscore{0.59}, 7 brothers\subscriptscore{0.71}, 9 of the Osmond siblings\subscriptscore{0.82} & 
        \cmark Alan Osmond\subscriptscore{0.89}\newline
        \cmark Wayne Osmond's\subscriptscore{0.72}\newline
        \cmark Merrill Osmond\subscriptscore{0.64}\newline
        \cmark Donny Osmond's\subscriptscore{0.64}
        % Wayne, Merrill, Donny, Marie, Jay
        \\\hline
        
        & how many wives did king solomon have 
        % \newline KG-answerable (2) 
        & 700 &
        \textbf{CNP$_\textit{rep}$:} 700 wives\subscriptscore{0.97} \newline
        \textbf{Synonyms:} 500 wives\subscriptscore{0.54}, seven hundred wives\subscriptscore{0.9} \newline
        \textbf{Subgroups:} three of his wives\subscriptscore{0.96}, three children\subscriptscore{0.86} 
        % 5 wives\subscriptscore{0.69}, 2\subscriptscore{0.64}, twice\subscriptscore{0.98} \newline
        \textbf{Incomparables:} 700 hundred wives\subscriptscore{0.85}, 1,000\subscriptscore{0.98}&
        % two men\subscriptscore{0.66} 
        \cmark Moti Maris\subscriptscore{0.84} \newline
        \textbf{x} Memphis\subscriptscore{0.62}
        % -
        \\\hline
        % & how many hof did kobe play with \newline Snippet-answerable \newline difficult question & 5 &
        % \textbf{CNP$_\textit{rep}$:} 5 times\subscriptscore{0.74} \newline
        % \textbf{Synonyms:} five NBA championships\subscriptscore{0.65}&
        % \\\hline
        % & \multicolumn{4}{c|}{\sg{Include ``inactive volcanoes in hawaii''}}
        
        & how many inactive volcanoes are in hawaii
        & 5 &
        \textbf{CNP$_\textit{rep}$:} five active volcanoes\subscriptscore{0.87} \newline
        \textbf{Synonyms:} four active volcanoes\subscriptscore{0.85}, five separate volcanoes\subscriptscore{0.73} \newline
        \textbf{Incomparables:} 169 potentially active volcanoes\subscriptscore{0.8}
        &
        \cmark Diamond Head\subscriptscore{0.95}\newline
        \cmark Mauna Kea\subscriptscore{0.41}\newline
        \cmark Haleakala\subscriptscore{0}\newline
        \\\hline
    \end{tabular}
    \end{adjustbox}
\end{table*}

\subsection{Case Study}
We pick up some challenging and interesting count queries which bring out the complexities of the problem and also showcase the capabilities of our system. The queries are collected in Table~\ref{tab:full_example}. 
% We make live queries, where the snippets are the the top-50 snippets returned by Bing at the time of submission.\footnote{queries made in June 2022.} 

The first query is our running example \textit{``how many songs did john lennon write for the beatles''} where the information need is for songs by Lennon with an additional condition that these are for the Beatles band. The complexity of this question comes through the snippets which indicate that other band members (George Harrison, and Paul Mc-Cartney) also wrote songs and that there are songs co-written by the band members.

CoQEx returns 73 as the answer inference and count contextualizations i) ``22 songs'' which belongs to \textit{Subgroups} category, since it comes from a snippet talking about \textit{lead guitarist george harrison wrote 22 songs}, ii) ``61 songs'' classified as a synonym comes from a competing source which says that ``Lennon wrote 61 songs credited to Lennon-McCartney all by himself'' iii) ``229 songs'', classified as \textit{Incomparable}, comes from a snippet about all the songs The Beatles as a band has written. 

Finding instances are much harder, with the top-5 instances being false positives (names of the band members) with composition credits varying vastly across songs and albums. CoQEx identifies one joint composition ``Maggie Mae'' and one album ``A Hard Day’s Night'' whose title track and majority of the album songs are written by Lennon, but is not very confident, scoring it very low (0.01).

The second query, \textit{``how many main islands in hawaii''}, is looking specifically for the \textit{main} Hawaiian islands. The CNPs returned by CoQEx, express the different interpretations of ``main'' as being more popular (``major'') or being ordered in terms of size (``largest''). CoQEx is also able to corroborate this with correct instances. 

The third query, \textit{``how many languages are spoken in indonesia''} seems relatively simple with a popular entity ``Indonesia'', well-defined predicate ``spoken in'' and an answer type ``language'', but is a great example of high variance answers. The presence of the modifier ``estimated'' and multiple close numbers (653, 700, 709) in the CNPs highlight the fact that it may not be possible to have one true answer. 

The fourth query, \textit{``how many osmund brothers are still alive''} is a query from the CoQuAD dataset, where instances in the snippets are more prevalent than counts. The CNPs ``9 siblings'' counts all brothers (8) and a sister, and ``7 brothers'' CNP belongs to snippet of the form ``Melvin Wayne Osmund has 7 brothers'', where the eighth brother is instantiated. CoQEx gets correct instances except for ``Marie'' who is the Osmund sister. Since all of the Osmund siblings were famous musicians, they pop up across relevant snippets.

The fifth query, \textit{``how many wives did king solomon have''} is interesting since KGs have two instances of Solomon's wives, which would lead a user to believe that 2 is the correct answer. However, multiple snippets confirm that the number is 700 and also provides a relevant instance ``Moti Maris'' which is not present in the current KGs.

The sixth query, \textit{``how many inactive volcanoes in hawaii''} is another interesting query which highlights the misunderstanding of document retrievers of \textit{inactive} volcanoes as \textit{active}, since all snippets returned deal with active volcano counts. Here, the instances are important, since those returned fall in the dormant or extinct categories of volcanoes.

Preliminary access to the system demonstrator is available at \url{https://nlcounqer.mpi-inf.mpg.de/}, where users can query pre-fetched snippets used in the CoQuAD dataset or make limited live queries dealing with counts.  

% \subsection{Demonstrator}
% \sg{Description and figures go here}

\section{Conclusion and Future Work}
% Explanatory evidence improve prediction comprehension through CNPs when there exist similar competing answers attributed to specificity of qualifiers (eight gold medals vs eight Olympic golds). TIme is also an important aspect which we leave for future work. 

% GoogleSDA may improve over time on popular searches, but 
We address the gap in distribution-aware prediction, assimilating semantic qualifiers from web contents and providing explanations through instances for the class of count queries. We systematically analysed count queries, their prevalence, structure and how current state-of-the-art answer them. 
We provide a thorough analysis of our system components and how {\mysystem} compares to the baselines in different query settings. A thorough discussion on tackled and open challenges are provided in the discussion section (Section~\ref{sec:discussion}) with observations and case studies. 

Improving explanation by instances has a major scope for improvement, by incorporating KB knowledge (for improving precision) or scraping list pages from search results (for improving recall). In Section~\ref{sec:results} we indicate ad-hoc mechanism to identify existing contextualizations in present KB-QA systems. This can be further expanded independently or CNPs from text could be useful in identifying relevant semantic qualifiers in the KB.  
% We can therefore still test the limits and improve the performance of the systems by querying on challenging qualifiers and less popular topics. 
% We presented {\mysystem} for answering count queries with explanatory evidence from multiple text segments, via inference, contextualization and explanation. Experiments show that {\mysystem} significantly outperforms deployed QA systems in precision and recall metrics. 

To foster further research, we release all datasets\footnote{\small\url{https://github.com/ghoshs/CoQEx}} 
% \footnote{\small\url{https://tinyurl.com/countqueryappendix}}
and provide access to the system demonstrator\footnote{https://nlcounqer.mpi-inf.mpg.de/} to make queries and see results through an interactive interface. 
% A web demonstrator is also deployed at \textit{[anonymized]}.

% %% Loading bibliography style file
\bibliographystyle{model1-num-names}

\bibliography{nlcounqer_refs}
\end{document}